\documentclass[usenatbib,fleqn]{mnras}
\pdfoutput=1
\usepackage{amsmath,amssymb}
\usepackage{graphics}
\usepackage{delarray}
\usepackage{graphicx}
\usepackage[table]{xcolor}
\usepackage{url}
\usepackage{rotating}
\usepackage{lscape}
\usepackage[T1]{fontenc}
\usepackage{ae,aecompl}

\newcommand{\fesc}{\rm f_{esc,\star}}

\newcommand{\zreion}{\rm z_{r}}

\newcommand{\Lbox}{\rm L_{box}}

\newcommand{\hmpc}{$\rm{h^{-1}Mpc}$ }
\newcommand{\chmpc}{$\rm{h^{-1}cMpc}$ }

\newcommand{\rev}[1]{\noindent{}}

\newcommand{\Msun}{\ensuremath{\mathrm{M}_{\odot}} }
\newcommand{\Msunnospace}{\ensuremath{\mathrm{M}_{\odot}}}

\newcommand{\nicefrac}[2]{\leavevmode\kern.1em
            \raise.5ex\hbox{\the\scriptfont0 #1}\kern-.1em
      /\kern-.15em\lower.25ex\hbox{\the\scriptfont0 #2}}

\newcommand{\hMpc}{{ \textit{h}$^{-1}$~Mpc}}
\newcommand{\hmsun}{{\, h^{-1}\rm~M}_\odot}
\newcommand{\music}{{\sc Music}}

\title[Cosmic Dawn II]{Cosmic Dawn II (CoDa II): a new radiation-hydrodynamics simulation 
of the self-consistent coupling of galaxy formation and reionization}

\author[P. Ocvirk et al.]{
  Pierre Ocvirk$^{1}$,
  Dominique Aubert$^{1}$,
  Jenny G. Sorce$^{1,2,3}$,
  Paul R. Shapiro$^{4}$,
  \newauthor
  Nicolas Deparis$^{1}$,
  Taha Dawoodbhoy$^{4}$,
  Joseph Lewis$^{1}$,
  Romain Teyssier$^{5}$,
  \newauthor
  Gustavo Yepes$^{6,7}$,
  Stefan Gottl\"{o}ber$^{3}$,
  Kyungjin Ahn$^{8}$,
  Ilian T. Iliev$^{9}$,
  \newauthor
  Yehuda Hoffman$^{10}$.
\\
$^1$Universit\'e de Strasbourg, CNRS, Observatoire astronomique de Strasbourg, UMR 7550, F-67000 Strasbourg, France\\
$^2$Universit\'e Lyon 1, ENS de Lyon, CNRS, Centre de Recherche Astrophysique de Lyon UMR5574, F-69230, Saint-Genis-Laval, France\\
$^3$Leibniz-Institut f\"{u}r Astrophysik Potsdam (AIP), An der Sternwarte 16, D-14482 Potsdam, Germany\\
$^4$Department of Astronomy, University Texas, Austin, TX 78712-1083, USA\\
$^5$Institute for Theoretical Physics, University of Zurich, Winterthurerstrasse 190, CH-8057 Z\"urich, Switzerland\\
$^6$Departamento de F\'{\i}sica Te\'orica M-8, Universidad Aut\'onoma de  Madrid, Cantoblanco 28049, Madrid, Spain\\
$^7$Centro de Investigaci\'{o}n Avanzada en F\'{\i}sica Fundamental (CIAFF), Facultad de Ciencias, Universidad Aut\'{o}noma de Madrid,\\
Cantoblanco  28049 Madrid, Spain\\
$^{8}$Chosun University, 375 Seosuk-dong, Dong-gu, Gwangjiu 501-759, Korea\\
$^9$Astronomy Center, Department of Physics \& Astronomy, Pevensey II Building, University of Sussex, Falmer, Brighton BN1 9QH, United Kingdom\\
$^{10}$Racah Institute of Physics, Hebrew University, Jerusalem 91904, Israel
}
\date{Accepted XXX. Received YYY; in original form ZZZ}

\pubyear{2018}

\hypersetup{draft}
\begin{document}
\label{firstpage}
\pagerange{\pageref{firstpage}--\pageref{lastpage}}
\maketitle

\begin{abstract}
Cosmic Dawn II (CoDa II) is a new, fully-coupled radiation-hydrodynamics simulation of cosmic reionization and galaxy formation and their mutual impact, to redshift $z < 6$. With $4096^3$ particles and cells in a 94 Mpc box, it is large enough to model global reionization and its feedback on galaxy formation while resolving all haloes above $10^8 \Msun$.  Using the same hybrid CPU-GPU code RAMSES-CUDATON as CoDa I in \cite{codaI}, CoDa II modified and re-calibrated the subgrid star-formation algorithm, making reionization end earlier, at $z \gtrsim 6$, thereby better matching the observations of intergalactic Lyman-alpha opacity from quasar spectra and electron-scattering optical depth from cosmic microwave background fluctuations. 
CoDa II predicts a UV continuum luminosity function in good agreement with observations of high-z galaxies, especially at $z = 6$.
As in CoDa I, reionization feedback suppresses star formation in haloes below $\sim 2 \times 10^9$ \Msun, though suppression here is less severe, a possible consequence of modifying the star-formation algorithm. Suppression
is environment-dependent, occurring earlier (later) in overdense (underdense) regions, in response to their local reionization times.
Using a constrained realization of $\Lambda$CDM constructed from galaxy survey data to reproduce the large-scale structure and major objects of the present-day Local Universe, CoDa II serves to model both \emph{global} and \emph{local} reionization. In CoDa II, the Milky Way and M31 appear as individual islands of reionization, i.e. they were not reionized by the progenitor of the Virgo cluster, nor by nearby groups, nor by each other.

\end{abstract}



%




\section{Introduction}

The first billion years of the Universe is a key period for the evolution of the intergalactic medium (hereafter, "IGM"), and the formation of galaxies. As the first stars form, their UV photons propagate, carving ionised regions which ultimately percolate, marking the end of the epoch of reionization (hereafter, "EoR"). The EoR is considered as the next observational frontier, both for 21cm radio telescopes such as LOFAR, MWA, SKA, NenuFAR, and infrared or near-infrared missions such as JWST. The recent tentative detection of a 21cm from the dark ages by EDGES  \citep{edges2018} has stirred the community and prompted numerous teams to try to confirm the signal with other instruments.

Meanwhile, a new breed of numerical simulations of galaxy formation has appeared, which aims at accounting for the rich physics, in particular the impact of the ionising radiation, necessary to describe 
the EoR\footnote{See \cite{dayal2018} for a review of recent advances in numerical simulations of galaxy formation and reionization.}. Faithful modelling of reionization, though, requires us to be able
to describe simultaneously the coupled multi-scale problem of global reionization and individual
galaxy formation, with gravity, hydrodynamics and radiative transfer. 
To represent the complex geometry of reionization in a statistically meaningful way requires a
a comoving computational domain as large as $\sim (100$ Mpc$)^3$ e.g. \cite{iliev2006,iliev2014}.
In order to follow the millions of galaxies in this volume over the full mass range of 
galactic haloes contributing to reionization\footnote{
the so-called ``atomic-cooling haloes" (henceforth, ``ACHs"), those with virial temperatures
above $\sim 10^4$ K, corresponding to halo masses above $\sim 10^8$ \Msun}, while modelling
the impact of reionization on these individual galaxies and the IGM,  
a physical resolution of a few kpc over the entire volume is needed.

The EoR is also important for its impact on low-mass galaxy formation.  The baryonic gas content of low-mass galaxies is
depressed by the hydrodynamical backreaction of their interstellar and surrounding intergalactic media, due to the 
photoheating that accompanies reionization by the rising ionizing UV background \citep{shapiro1994,gnedin2000,hoeft2006}.
This, in turn, suppresses their star formation rates. 
The resulting ``fossil galaxies" may offer a credible solution to 
the missing satellite problem as formulated in the Local Group 
\citep[see, e.g.,][and references, therein]{BBK2017}.

The faintest galaxies known, potentially susceptible to such radiative suppression of their gas and star formation, 
can only be detected nearby, in the Local Group. Therefore, to make direct 
comparisons possible between simulations and this observationally-accessible sample, 
it is necessary to start from initial conditions preselected to produce the observed galaxies and
large-scale structure of the Local Universe.

In the past few years, we have made significant progress in this direction, in our
project of large-scale simulation of cosmic reionization and galaxy formation based 
upon fully-coupled radiation-hydrodynamics, the Cosmic Dawn (``CoDa") simulation project.
The first of the CoDa simulations (hereafter, referred to as ``CoDa I"), described in
\cite{codaI} (hereafter, O16), with further results presented in \cite{dawoodbhoy2018},
was based upon the massively-parallel, hybrid CPU-GPU code RAMSES-CUDATON and ran
for 11 days on the Titan supercomputer at the Oak Ridge Leadership Computational Facility 
(OLCF), with 8092 GPUs and 8092 CPUs, 
utilizing $4096^3$ particles and cells in a volume 94 comoving
Mpc on a side.  The second simulation, CoDa I-AMR, described in \cite{codadom}, 
was based upon the massively-parallel, hybrid CPU-GPU code EMMA, 
with Adaptive Mesh Refinement (AMR), which also ran on Titan, 
with 4096 GPUs and 32768 CPUs, utilizing $2048^3$ particles, 
starting from a uniform grid of $2048^3$ unrefined cells from which AMR 
increased the resolution locally, by up to a factor of 8, depending on 
the local overdensity,  leading ultimately to 
18 billion cells after refinement, in the same size box, 
starting from the same initial conditions (except coarsened to the initial, unrefined 
grid of CoDa I-AMR).  

These CoDa simulations share four principal characteristics
that make them special and a first of their kind, as follows.   
Their resolution is high enough to track the formation of the 
mass range of galactic haloes above $10^8 \Msun$ believed to dominate reionization 
and the back-reaction of reionization on their evolution as sources.  
Their volume is large enough to model the mean history of reionization and the 
hydrogen-ionizing UV background, along with the inhomogeneity of reionization in time and space.  
They are based upon a constrained realization of the Gaussian random initial conditions
for $\Lambda$CDM, derived from galaxy observations of the Local Universe, which are designed
to reproduce the large-scale structure and familiar features observed in our Local Universe
today when evolved to the present. By starting from this ``constrained
realization" of initial conditions which reproduces a number of selected, familiar 
features of the Local Universe in a volume centered on the Local Group, 
such as the Milky Way, M31, and the Fornax and Virgo clusters, we are able
to use these simulations to model both \emph{global} and 
\emph{local} reionization, simultaneously, including the impact of reionization on
the formation of galaxies in the Local Group.  
Finally, by fully-coupling the hydrodynamics to the radiative transfer of ionizing
radiation, we move beyond simulations which adopt a pre-computed, uniform UV background,
by computing the rise of the UV background, inhomogeneous in space and time, 
and its location-specific impact on galaxies and their emissivity, together,
self-consistently.

In this paper, we present the Cosmic Dawn II (``CoDa II") simulation, which aims to improve upon
our previous two simulations in several aspects. In CoDa I (O16), the end of global reionization was
somewhat later than observed, at $z \lesssim 5$, a consequence of adopting a slightly lower efficiency
for the subgrid star formation algorithm than necessary to end reionization by $z = 6$, despite
our effort to calibrate this efficiency by a series of small-box simulations intended to predict
the outcome for the large box.  For CoDa II, the star formation algorithm parameters were re-calibrated and
the algorithm was also modified, to ensure that reionization ended earlier than CoDa I, by $z = 6$,
in better agreement with the observations. While CoDa I-AMR \citep{codadom} was also successfully re-calibrated to
end reionization earlier than did CoDa I, both CoDa I and CoDa I-AMR started from the same
generation of ``constrained realization" initial conditions developed by the \emph{Constrained Local UniversE Simulations}
(``CLUES") project which have since been updated and improved, including for instance more recent observational data.
Those earlier initial conditions resulted in a Virgo replica less 
massive than observed, and were derived for a background universe with cosmological 
parameters for a WMAP5 cosmology \citep{hinshaw2009}. The new initial conditions for CoDa II now feature a more realistic Virgo 
replica, and use a more recent set of cosmological parameters determined by the $\emph{Planck}$
collaboration \citep{2014A&A...571A..16P}.

The main goal of this first CoDa II paper 
is to introduce the simulation and compare the results with current observational constraints 
on the global EoR and the high-redshift galaxies that caused it, 
such as the evolution of the cosmic neutral and ionized fractions, 
the mean intensity of the cosmic ionizing UV background, the cosmic star formation rate density, 
the Thomson-scattering optical depth through the intergalactic medium measured by 
cosmic microwave background fluctuation measurements,
and the high-redshift galactic UV luminosity functions.  
In Sec. \ref{s:methodology}, we describe the
RAMSES-CUDATON code on which the CoDa II simulation is based,
and the simulation set-up, including the new ``constrained realization" 
initial conditions used here. Our results are
presented in Sec. \ref{s:results}, including the global
history of reionization compared with observational constraints, 
the star formation rates of galaxies and the net star formation
rate density of the universe, and the UV continuum luminosity function of high-z galaxies during the EoR. 
We use the simulation to show how the star formation rate in low-mass haloes is affected by reionization 
and the environmental dependence of this process. Finally, we investigate the reionization history of the Local Group, and then finish with a short summary in
Sec. \ref{s:conclusions}.

\section{Methodology}
\label{s:methodology}

The Cosmic Dawn simulation uses the fully coupled radiation hydrodynamics code RAMSES-CUDATON described in O16. This section outlines the basics of the code and the differences with CoDa I, as well as its deployment. The reader is referred to O16 for further detail. For quick reference, the parameters of the simulation are summarized in Table \ref{t:sum}.

\begin{figure*}
  {\includegraphics[width=1.\linewidth,clip]{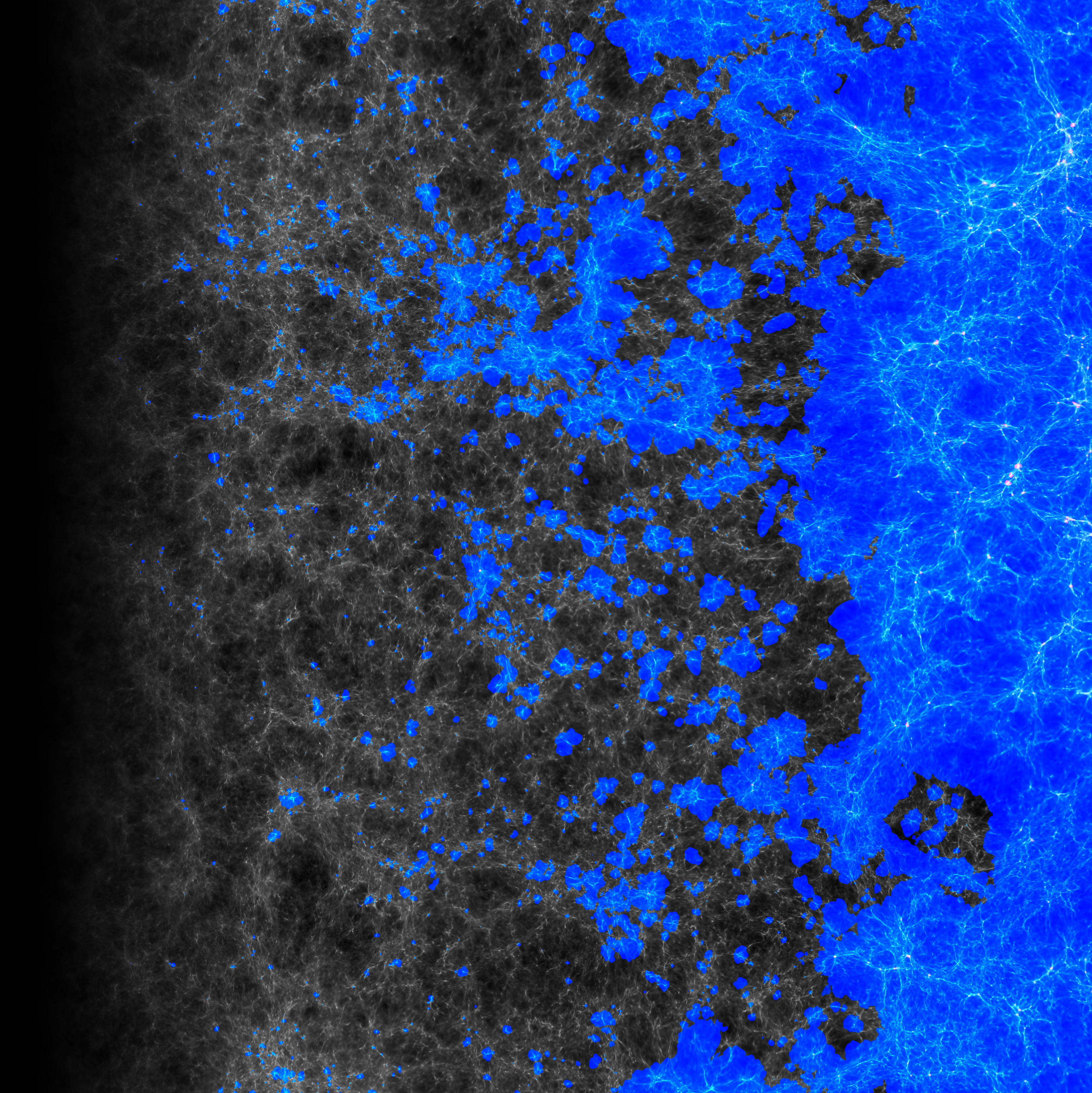}}
   \caption{Illustration of the time history of reionization in the CoDa II simulation: a pseudo-light-cone image.
A single, planar (x,y)-slice through the full $(94 cMpc)^3$ simulation cube was chosen for display, and a sequence 
of $\sim$ 1024 closely-spaced time-slices of the $4096^2$ grid cells in that plane was outputted, to sample the 
time-evolution from redshift z$=150$ to z$=5.8$.  
The image is made by the concatenation of vertical stripes, 4-cells wide, each stripe selected from
another time-slice in the sequence, to show the time evolution as time increases to the right along 
the x-axis, from redshift z$=150$ (left) to z$=5.8$ (right).
Blue regions are photo-heated, 
while small, bright red regions (which can be seen more clearly by zooming the high-res figure)
correspond to regions heated by supernovae feedback and accretion shocks.
The green color, on the other hand, denotes regions where ionization is ongoing and incomplete, and temperature has not yet risen to the $\sim 10^4$ K typical of fully ionized regions. Brightness indicates the gas density contrast.}
\label{f:bigmap}
\end{figure*}

\begin{table}
\begin{tabular}{lr}
\hline
\multicolumn{2}{c}{Cosmology (Planck14)} \\
\hline
Dark energy density $\Omega_{\Lambda}$  & 0.693 \\
Matter density $\Omega_{\rm{m}}$  & 0.307 \\
Baryonic matter density $\Omega_{\rm{b}}$  & 0.048 \\
Hubble constant $h={\rm H}_0/(100 \, {\rm km/s})$  & 0.677 \\
Power spectrum & \\
$\,$ Normalization $\sigma_8$       &       0.8288 \\
$\,$ Index $n$  & 0.963 \\
\\
\hline
\multicolumn{2}{c}{Setup} \\
\hline
Number of nodes (GPUs, cores used)  & 16384 (16384, 65536) \\
Grid size   &	$4096^{3}$ \\
Comoving box size $\Lbox$ &   94.44 Mpc (64 \hmpc)  \\
Grid cells per node  & 128x128x64 \\
Comoving force resolution dx & 23.06 kpc \\
Physical force resolution at z=6 & 3.3 kpc \\
DM particle number $N_{DM}$   &	$4096^3$ \\
DM particle mass $M_{DM}$	 & 4.07 x $10^5$ \Msun \\
Average cell gas mass   & 0.75 x $10^5$ \Msun \\
Initial redshift $z_{start}$  & 150 \\
End redshift $z_{end}$   & 5.8 \\
\\
\hline
\multicolumn{2}{c}{Star formation}\\
\hline
Density threshold $\delta_{\star}$  & $50 \, \langle \rho_{\rm gas} \rangle$ \\
Efficiency $\epsilon_{\star}$  & $0.02$  \\
Stellar particle birth mass $M_{\star}$  & 11732 \Msun \\
\\
\hline
\multicolumn{2}{c}{Feedback}\\
\hline
Massive star lifetime $t_{\star}$  & 10 Myr \\
\hline
\multicolumn{2}{c}{Supernova}\\
\hline
Mass fraction $\eta_{SN}$  & 10\% \\
Energy $E_{SN}$  &  $10^{51}$ erg \\
\hline
\multicolumn{2}{c}{Radiation} \\
\hline
Stellar ionizing emissivity  & 4.32$\times 10^{46}$ ph/s/\Msun \\
Stellar particle escape fraction $\fesc$  & 0.42 \\
Effective photon energy   & 20.28 eV \\
Effective HI cross-section $\sigma_E$ & 2.493 x $10^{-22}$m$^2$ \\ 
Speed of light $c$  & 299 792 458 m/s \\
\end{tabular}
\caption{Cosmic Dawn II simulation parameters summary}
\label{t:sum}
\end{table}

\subsection{RAMSES-CUDATON}
\label{s:ramsescudaton}

Since, the code is already described in detail in O16, we here only give a brief summary of it and highlight the differences in implementation / parameter choices between CoDa I and CoDa II.
RAMSES-CUDATON results from the coupling of RAMSES \citep{teyssier02} and ATON \citep{aubert2008}. RAMSES handles gravity, hydrodynamics and star formation and its mechanical feedback, while ATON handles photon propagation, hydrogen-photon thermochemistry, and computes the cooling terms. Finally, CUDATON results from the porting of ATON to CUDA \citep{aubert2010} to enable it to take advantage of hybrid architectures featuring GPGPU (General Purpose Graphics Processing Unit). Thanks to this unique feature, we were able to perform CoDa II using the full speed of light, therefore circumventing potential artefacts arising with reduced speed of light approaches, highlighted by  \cite{deparis2018} and \cite{ocvirk2019}, or arising with infinite speed of light approaches in many ray-tracing codes. However, with this optimization, we are only able to use RAMSES in unigrid mode, i.e. without adaptive mesh refinement (hereafter, "AMR"). This is one of the shortcomings that the EMMA code \citep{aubert2015} and the RAMSES-RT code \citep{rosdahl2013} attempt to address.

\subsection{Differences with CoDa I}

\subsubsection{Star formation and calibration}

As in O16, we consider star
formation using a phenomenological approach. In each
cell with gas density larger than a gas overdensity $\delta_{\star}=50$, we spawn new star particles
at a rate given by
\begin{equation}
\dot{\rho}_{\star} = \epsilon_{\star} \frac{\rho_{gas}}{t_{ff}},
\label{sfr}
\end{equation}

\noindent where $t_{ff}=\sqrt{\frac{3 \pi}{32 G \rho}}$ is the free-fall time of the gaseous component and
$\epsilon_{\star}$= 0.02 is the star formation efficiency, instead of 0.01 in O16. The higher star formation efficiency of CoDa II is intended to achieve the end of reionization by redshift 6, whereas CoDa I finished reionization by z=4.6.  In practice,
we calibrate this subgrid star formation efficiency by running a suite of reionization simulations in much smaller boxes
than our final production run, but at the same grid and particle resolution as the latter, and then test the 
best-choice efficiency parameter by another suite of simulations of increasing box size.  We adjusted the value of
$\epsilon_{\star}$ so as to achieve a good agreement with the evolution of the cosmic star formation rate density
inferred from observations of high-redshift galaxies (discussed in Section 3.3).

Unlike the first Cosmic Dawn simulation reported in O16, we do not require the cell temperature to be lower than $T_{\star}=2 \times 10^4$ K in order to form stars: all cells, no matter their temperature, are eligible to forming stars above the density threshold. This is to account for the fact that, at higher resolution, such ionized cells may still host cold, neutral regions, which may still form stars. We will see in Sec. \ref{s:results} that this choice has an impact on the strength of the radiative suppression of star formation.

The star particle mass at birth depends on the cell gas density, but is always a multiple of a fixed elementary mass $M_{\star}^{birth}$, chosen to be a small fraction of the baryonic mass resolution. In this framework, with the box size and resolution of CoDa (see Sec. \ref{s:setup}), we have $M_{\star}^{birth}=11732$ \Msunnospace. This mass is small enough to sample adequately the star formation history of even low mass galaxies, and still large enough to mitigate stochastic variations in the number of massive stars per star particle.
It is larger than the elementary mass used in CoDa I, which had $M_{\star}^{birth}=3194$ \Msunnospace. This change reduces the number of star particles in the simulation, which in turn reduces their computational cost, in memory and processing.

\subsubsection{Ionizing source model}

The radiative transfer considered here is of H ionizing radiation released by stellar sources associated with the star particles
described in the previous section, subject to absorption by H I bound-free opacity. As with our subgrid star-formation
algorithm, the rate of release of ionizing photons by star particles into the grid cells in which they are located must
be parameterized, so as to represent the absorption taking place at the sub-pc scales of the interstellar molecular clouds where stars are formed inside each galaxy.

Hence, we must add to our subgrid star formation algorithm described above a parameterized ionizing photon efficiency
(henceforth, ``IPE"), the number of ionizing photons released per unit stellar baryon per unit time, into the host
grid cell of each star particle.   This IPE is based on an assumed stellar IMF and, since star particles are assumed
to form embedded in high-density molecular clouds which are unresolved by our grid cells, we also adopt a subgrid
stellar-birthplace escape fraction.  
We define this IPE, therefore, as $\xi_\textsc{ipe} \equiv f_{\rm{esc,\star}} \xi_{\text{ph,}\textsc{imf}}$, where 
$f_{\rm{esc,\star}}$ is the stellar-birthplace escape fraction and $ \xi_{\text{ph,}\textsc{imf}}$      
is the number of ionizing photons emitted per Myr per stellar baryon.  The subgrid stellar-birthplace escape fraction
$f_{\rm{esc,\star}}$ should  not be confused with the \emph{galactic} escape fraction, which is not a subgrid quantity
put in ``by hand" but is rather the self-consistent outcome of our radiative transfer module CUDATON between grid
cells, for all the cells associated with a given galactic halo.   There is no need for us to adopt a galactic escape fraction, therefore.  Our RT calculation finds the actual time-varying escape fraction for each galactic
halo, which differs from one galaxy to the next.  

Each stellar particle is considered to radiate for one massive star lifetime $t_{\star}=10$ Myr, 
after which the massive stars die (triggering a supernova explosion) and the particle becomes dark in the H-ionizing UV.
We adopted an emissivity $ \xi_{\text{ph,}\textsc{imf}} = 1140$ ionizing photons/Myr per stellar baryon. 
This is similar to the ionizing emissivity of a BPASS model \citep{bpass21} for a Z $=0.001$ binary population 
with Kroupa initial mass function \cite{kroupa2001}, as presented in \cite{rosdahl2018},  
averaged over the first 10 Myr.
Although we do not follow chemical enrichment in CoDa II, \cite{pawlik2017} showed that the metallicity in gas
at the average density of the ISM is indeed close to Z $=0.001$ during the epoch of reionization, which validates the choice of this metallicity.
We used a mono-frequency treatment of the radiation with an effective frequency of 20.28 eV,
%
as in \cite{baek2009}. 

Finally, we calibrated the stellar-particle escape fraction  $f_{\rm{esc,\star}}$ by adjusting the value
in our set of smaller-box simulations as mentioned above, so as to obtain a reionization redshift close to 
$z = 6$, which led us to adopt a value of $\fesc=0.42$, close to the value $\fesc=0.5$ used in O16.  We note that the 
$\it{net}$ escape fraction from the cell in which the star particle is located is less than or equal to 
this subgrid value adopted for the unresolved stellar-birthplace, since there is additional absorption of ionizing photons
inside the cell, by the bound-free opacity of the simulated gas in that cell, which depends upon that cell's 
neutral H fraction and gas density, as tracked self-consistently by RAMSES-CUDATON.  The over-all
$\it{galactic}$ escape fraction will be even smaller, since galactic halos typically involve multiple cells, 
and one cell can absorb the photons released in another.   Our radiative transfer module CUDATON explicitly
accounts for this.

As in Coda I, we neglect here the possible release of hard, ionizing UV radiation associated with 
a phase in the evolution of our simulated galaxies in which they host
Active Galactic Nuclei (hereafter, AGN).  Some AGN sources may arise during the epoch of reionization, 
e.g. in rare, massive proto-clusters \citep{dubois2012}.   However,  
they are very rare and have long been thought to be minor contributors to the cosmic budget 
of ionizing photons responsible for the 
reionization of hydrogen \citep[e.g.][]{shapiro87,giroux96,haardt2012,haardt2015}, 
although they could be important for explaining the line of sight variations of the properties of 
the Ly $\alpha$ forest just after reionization \citep{chardin2015a}.  The question of the
possible AGN contribution to reionization has recently been revisited following claims by 
\cite{2015A&A...578A..83G} that the observed AGN luminosity function at high redshift may have been 
underestimated \citep{madau15}.  Further studies
indicate that the AGN contribution to reionization must be subdominant or else would violate 
other observational constraints
\citep[e.g.][]{daloisio17,worseck16,onorbe17,mitra18,qin17}, 
while subsequent updates of the observed quasar luminosity function also conclude that the AGN contribution 
to reionization is very small \citep[e.g.][]{parsa18,2019MNRAS.488.1035K}.

\subsection{Simulation setup}
\label{s:setup}

\subsubsection{Initial conditions}
\label{s:ics}
Our initial conditions are a constrained realization of the $\Lambda$CDM universe, 
intended to reproduce the observed
features of our Local Universe in a box centered on the Milky Way if evolved to the present, 
while at the same time serving as a representative sample of
the universe-at-large with which to model global reionization.  We generated  specifically the initial conditions for this CoDa II simulation as part of the CLUES ({\it Constrained Local UniversE Simulations})
project. These initial conditions are 
constrained by observational data on the positions and velocities of galaxies 
in the Local Universe described in \citet{2013AJ....146...86T} (referred to as {\it Cosmicflows-2})
to result in a simulation that resembles the Local Universe within the framework of 
$\emph{Planck}$ cosmology \citep[$\Omega_m=0.307$, $\Omega_\Lambda=0.693$, $H_0=67.77$ km~s$^{-1}$~Mpc$^{-1}$, $\sigma_8$~=~0.829,][]{2014A&A...571A..16P}, 
in a comoving box 94 (i.e. 64~$h^{-1}$) cMpc on a side. 
It contains 4096$^3$ dark matter particles on a cubic-lattice with the same number of cells, 
which are also the grid cells for the gas and radiation. 
The mass of each dark matter particle is then 4.07$\times$10$^5$~M$_\odot$. 
The CoDa II initial conditions have the average universal density for this chosen cosmology. 

\citet{2016MNRAS.455.2078S} describes in detail the steps of the method used to build these 
constrained initial conditions. The main steps are  summarized as follows:
\begin{itemize}
\item We started with a set of data points, the radial peculiar velocities of observed galaxies,
i.e. radial peculiar velocities $v_r$ at discrete positions $r$ at ${z = 0}$.
Galaxies in the radial peculiar velocity catalog were then grouped \citep[e.g.][]{2015AJ....149..171T,2015AJ....149...54T} 
to produce a data set that traces the coherent large-scale velocity field,
with non-linear virial motions removed that would affect the reconstruction obtained with the linear method as shown by \citet{2017MNRAS.468.1812S,2017MNRAS.469.2859S}.
\item Biases inherent to any observational radial peculiar velocity catalog were minimized \citep{2015MNRAS.450.2644S}.
\item The three-dimensional peculiar velocity field and associated cosmic displacement field were then
reconstructed by applying the Wiener-filter (henceforth, WF) technique -- a linear minimum variance estimator 
\citep{1995ApJ...449..446Z,1999ApJ...520..413Z}
-- to the grouped and bias-minimized radial peculiar velocity constraints.
\item To account for the displacement of mass elements located at the positions of the
galaxies in the catalogue at ${z = 0}$, away from the positions of their Lagrangian patch in the early universe,
caused by the growth of structure over time, the positions of the constraints were relocated
to the positions of their progenitors using the Reverse Zel'dovich Approximation \citep{2013MNRAS.430..888D}. Additionally, noisy radial peculiar velocities are replaced by their WF 3D reconstructions \citep{2014MNRAS.437.3586S}. 
\item Linear density and velocity fields were then produced, constrained by the modified observational peculiar velocities 
combined with a random realization of $\Lambda$CDM to restore statistically the 
missing structures using the Constrained Realization technique \citep[][]{1991ApJ...380L...5H,1992ApJ...384..448H}. 
\item These density and velocity fields, extrapolated to ${z = 0}$ according to linear theory,
have to be rescaled to an actual starting redshift (${z_i = 150}$ for the CoDa II simulation) to build its constrained initial conditions, and the resolution was increased
\citep[\music,][]{2011MNRAS.415.2101H}. Increasing the resolution implies adding higher-frequency modes to the
initial conditions, which introduce random small-scale features. 
For this purpose, we created 200 random realizations of the initial conditions with these higher-frequency modes added,
and tried them out by dark-matter-only N-body simulations [using the GADGET-2 code \citep{springel2005}] evolved to $z = 0$,
in order to select that with the best match to the observed features of the Local Universe at $z=0$, 
including the Local Group\footnote{The Local Group belongs to the non-linear regime 
and is thus not directly constrained but is induced by the local environment \citep{2016MNRAS.458..900C}.}
(according to its mass, the separation between the Milky Way and M31, their mass ratio, and 
very nearby environment with M33 and Centaurus A counterparts). The production of other prominent 
features of the Local Universe, like the Virgo cluster (with close to the right mass at close to the right
position), was also used as a selection criterion.

The CoDa II initial conditions were thus chosen so as to reproduce familiar features
of the observed Local Universe within a volume 94 cMpc on a side, centered on the LG, as closely as 
possible. Towards this end, all 200 of the dark-matter-only N-body simulations from these 
200 constrained realizations (as described above)
were examined at $z=0$, and, amongst these, we selected a reduced set which best reproduced the Local Group,
then further selected amongst these to meet our additional requirements, as follows: 

\begin{figure*}
    \centering
    \includegraphics[width=.49\linewidth]{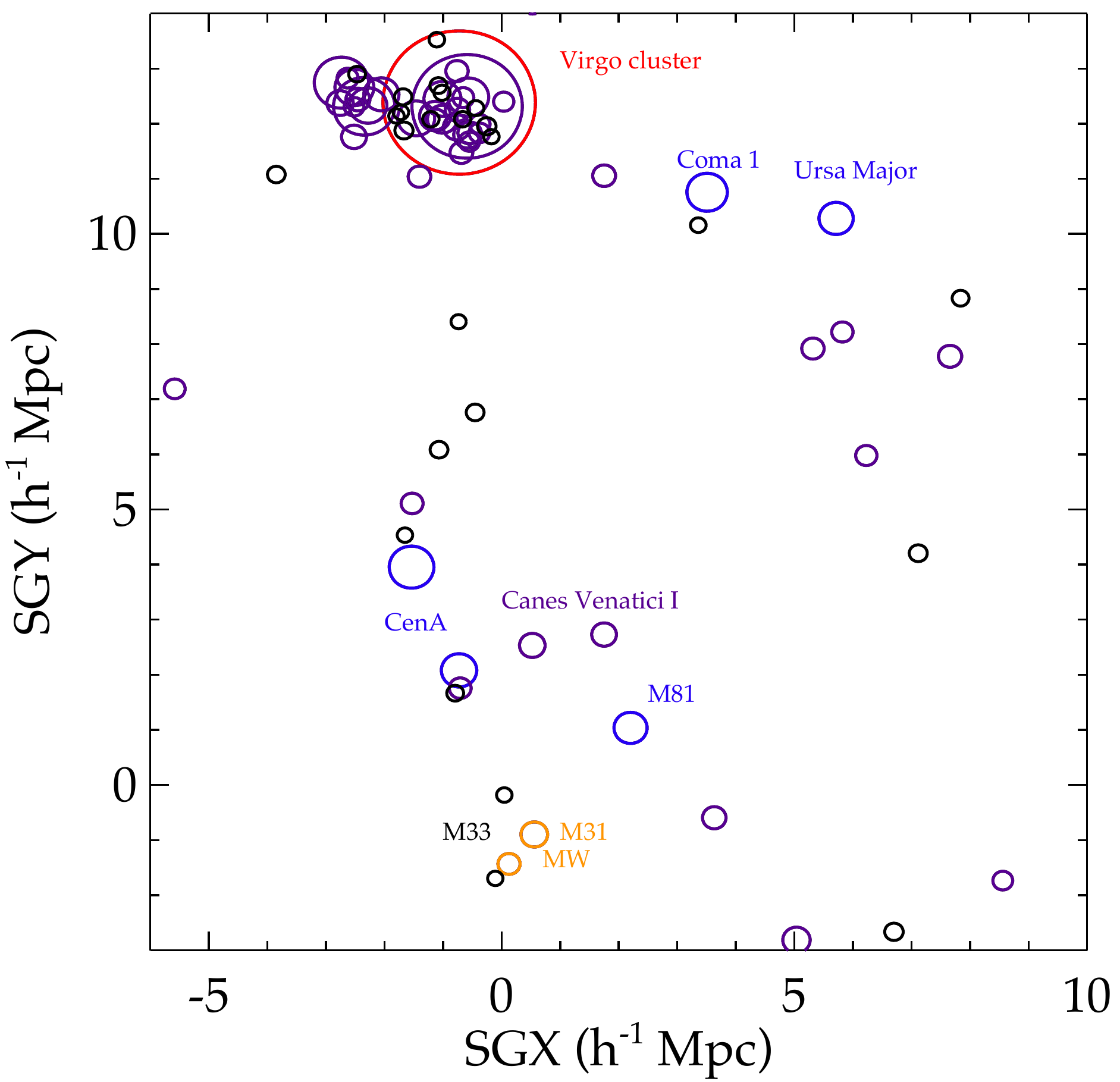}
    \includegraphics[width=.49\linewidth]{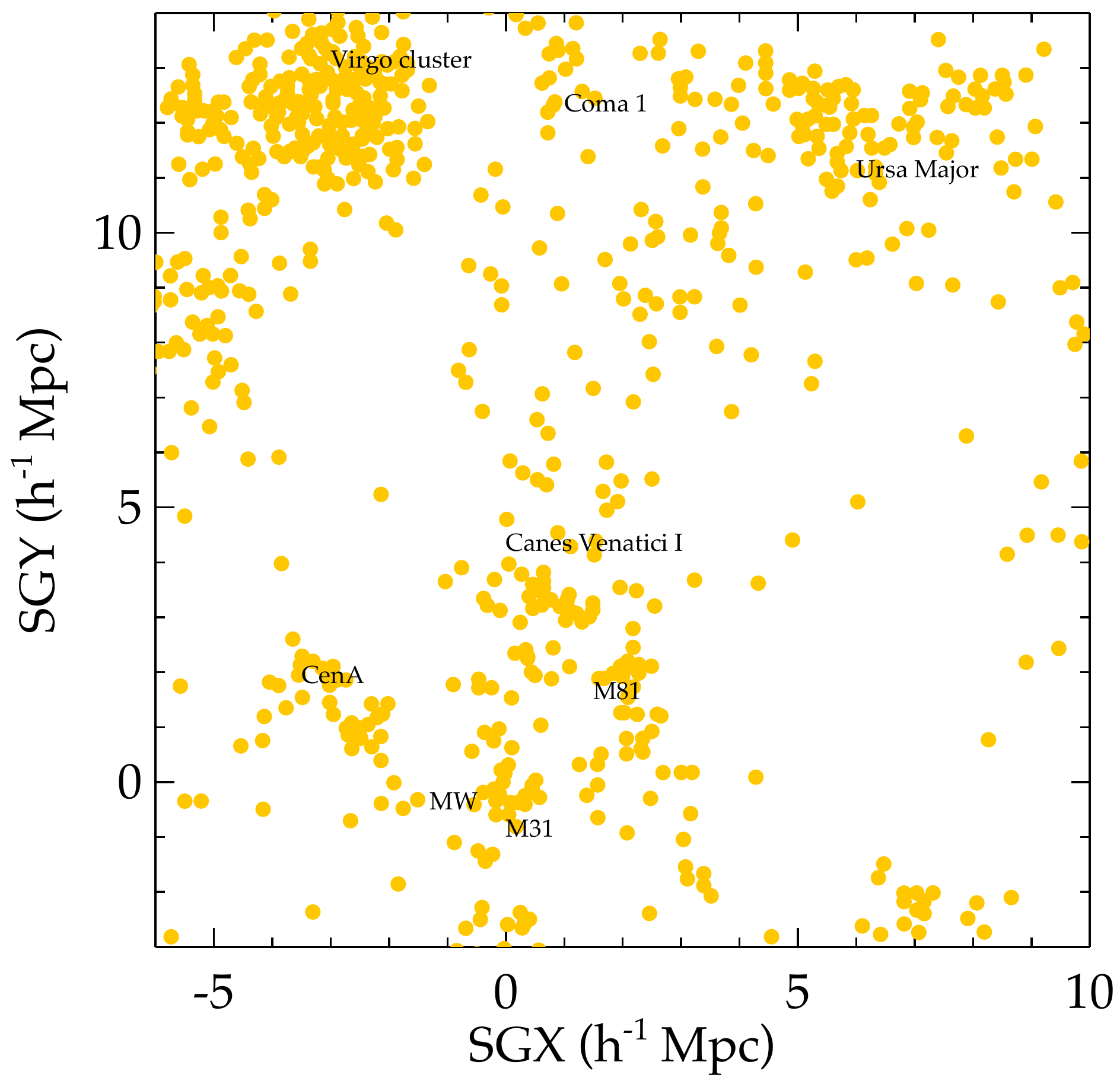}
    \caption{Comparison of the N-body halos with masses $M > 2\times 10^{11}$ M$_\odot$
    in CoDa II-DM2048 at z = 0 in the Supergalactic Plane,
    in a subregion 16 $h^{-1}$ Mpc on a side and 4 $h^{-1}$ Mpc thick (left panel)
    with the observed galaxies, groups and clusters of the nearby Local Universe for the same region, a thin slice
    centered on the supergalactic equator, 4 $h^{-1}$ Mpc thick (right panel),
    based upon the observed galaxies as plotted in Fig. 4 of \citet{courtois2013}, taken from the Extragalactic Distance Database ({\url{http://edd.ifa.hawaii.edu}}). Circles of radius equal to the virial
    radii of the simulated halos are shown on the left panel.  Features of the Local Universe and their analogs
    in the simulation are labelled. The correspondence between figure labels and Simbad (\url{http://simbad.u-strasbg.fr/simbad/}) names is given in Tab. 2.}
    \label{fig:codaii-dm2048-z0}
\end{figure*}

(i) A pair of galaxies is located at the center (by construction) with masses, M$_{200}$,
between $5.5\times10^{11}$ and $2\times10^{12} h^{-1}$ M$_\odot$.

(ii) There is no other halo more massive than $5.5\times10^{11} h^{-1}$ M$_\odot$
within a sphere of radius 2.5 $h^{-1}$ Mpc.

(iii) Their separation is smaller than 1.5 $h^{-1}$Mpc.

(iv) Their mass ratio is smaller than 2.

(v) They are located between 10 $h^{-1}$ and 14 $h^{-1}$ Mpc away from the
Virgo cluster replica.

(vi) There are halos that could stand for M33 and Centaurus A (by
far the most restrictive criterion).

Left with a dozen simulations with halo pairs, 
we selected the pair (and, thus, the simulation)
that satisfied both requirements, that of a small separation and a small
mass ratio: at z=0, the distance between the two halos of the pair is
0.85 $h^{-1}$ Mpc and their mass ratio is 1.2 (with masses of
$1.55\times10^{12}$ and $1.3\times10^{12} h^{-1}$ M$_\odot$).

These N-body simulations were at lower mass-resolution 
than our final CoDa II simulation but at a resolution high enough 
(at 512$^3$ particles, with particle mass 1.7$\times$10$^8~\hmsun$) 
to ensure that the resulting Local Group candidates would be stable to increasing the 
resolution still further when the same initial conditions were resimulated in our CoDa II production 
run \footnote{The resolution of each of these N-body simulations must indeed be large enough 
for their candidate Local Group to be affected only barely when 
resimulated with even higher resolution, when yet higher-frequency random modes are introduced
to the initial conditions, thereby adding even smaller-scale features.}. 
The resolution of the best initial conditions thereby selected was then 
increased still further, to 4096$^3$ particles and cells, adding additional modes at higher frequency, still with \music.
In order to provide a companion dark-matter-only N-body simulation to $z=0$ from these
constrained-realization initial conditions, we also coarsened the final, high-resolution CoDa II
initial conditions from $4096^3$ down to 2048$^3$, to run a GADGET-2 N-body simulation to $z=0$,
called CoDa II-DM2048\footnote{It corresponds to the simulation ESMDPL\_2048 from the MultiDark project. It is fully available at www.cosmosim.org}. The CoDa II-DM2048 simulation was then
used to make a final comparison of the simulated LG and other
features of the Local Universe by our CoDa II initial conditions with
the observed Local Universe at $z=0$.  A comparison of the observed Local Universe with the outcome
of simulating the CoDa II initial conditions to $z=0$ by CoDa II-DM2048, is shown in Figure~\ref{fig:codaii-dm2048-z0}. 
\end{itemize}

The advantage of these new initial conditions with respect to those used for the first generation of CoDa simulations, CoDa I and CoDa I-AMR, is a Large-Scale Structure that matches the local one down to the linear threshold ($\sim$3~\hMpc) on larger distances (the entirety of the box if not for the periodic boundary conditions) and with more accurate positions ($\sim$3-4~\hMpc). In particular, this simulation contains a Virgo cluster at the proper position and with a mass in better agreement with recent observational mass estimates \citep{2016MNRAS.460.2015S}. While the Virgo replica had a mass of 7$\times$10$^{13}~\hmsun$ in the CoDa I simulation, its mass is now 2.2$\times$10$^{14}~\hmsun$, i.e. the cluster is 3 times more massive\footnote{Higher masses in prefect agreement with observational estimates for Virgo have been obtained in realizations of $500$ \hMpc, but with the fixed $4096^3$ grid size used here, such a box size results in a spatial and mass resolution unsuitable for this study.}. These properties will be important for the planned follow-up projects.
 
Finally, the baryonic initial conditions at the initial redshift of $z = 150$
were generated assuming a uniform temperature equal to that of the CMB at that time, 
with identical gas and dark matter velocity fields. 
The initial value of the H ionized fraction was taken to be the homogeneous, post-recombination-era
freeze-out value at $z = 150$, as computed following standard recipes such as in RECFAST (Seager et al. 1999). 

\begin{table}
    \centering
    \begin{tabular}{|l|l|}
    \hline
    Name & Simbad name \\
    \hline
    Virgo cluster & Vir I \\
    Coma I & Coma I group \\
    Ursa Major & Ursa Major cluster \\
    Canes Venatici I & CVn group \\
    CenA & Cen A group \\
    M81 & M81 group \\
    \hline
    \end{tabular}
    \caption{Correspondence table between names used in the article and Simbad names.}
    \label{tab:my_label}
\end{table}

\subsubsection{Code deployment}
Like CoDa I (O16) and CoDa I-AMR \citep{codadom}, 
CoDa II was performed on the massively-hybrid CPU-GPU supercomputer Titan, at OLCF.
The code was deployed on 16384 Titan nodes and 16384 GPUs
(1 GPU per node), with each GPU coupled to 4 cores, for a total of 65536 cores. 
Each node hosted 4 MPI processes that each managed a volume of $64\times128\times128$ cells.

\subsection{Post-Processing: Friends-of-Friends halo catalogs}
\label{s:fof}
We used the massively-parallel Friends-of-Friends (hereafter, FoF) halo finder of \cite{roy2014} with a standard linking length of $b=0.2$ to detect dark matter haloes in the CoDa II simulation. As shown in O16, they are reliably detected down to $\sim 10^8$ \Msun.

\subsection{Online data publication}
We plan to make a subset of the data and higher level products publicly available through the VizieR database at CDS Strasbourg\footnote{\url{http://vizier.u-strasbg.fr/viz-bin/VizieR?-source=VI/146}}. The latter link will be active only after acceptation of the paper. The CoDa II catalogs will also be made available through the cosmosim database hosted by Leibniz Institut f\"ur Astrophysik Potsdam\footnote{\url{https://www.cosmosim.org/cms/simulations/cosmic-dawn/}}.

\section{Results}
\label{s:results}
\subsection{Global properties}
\label{s:glob}
\begin{figure*}
\begin{center}
\begin{tabular}{cc}
  {\includegraphics[height=7.8cm,clip]{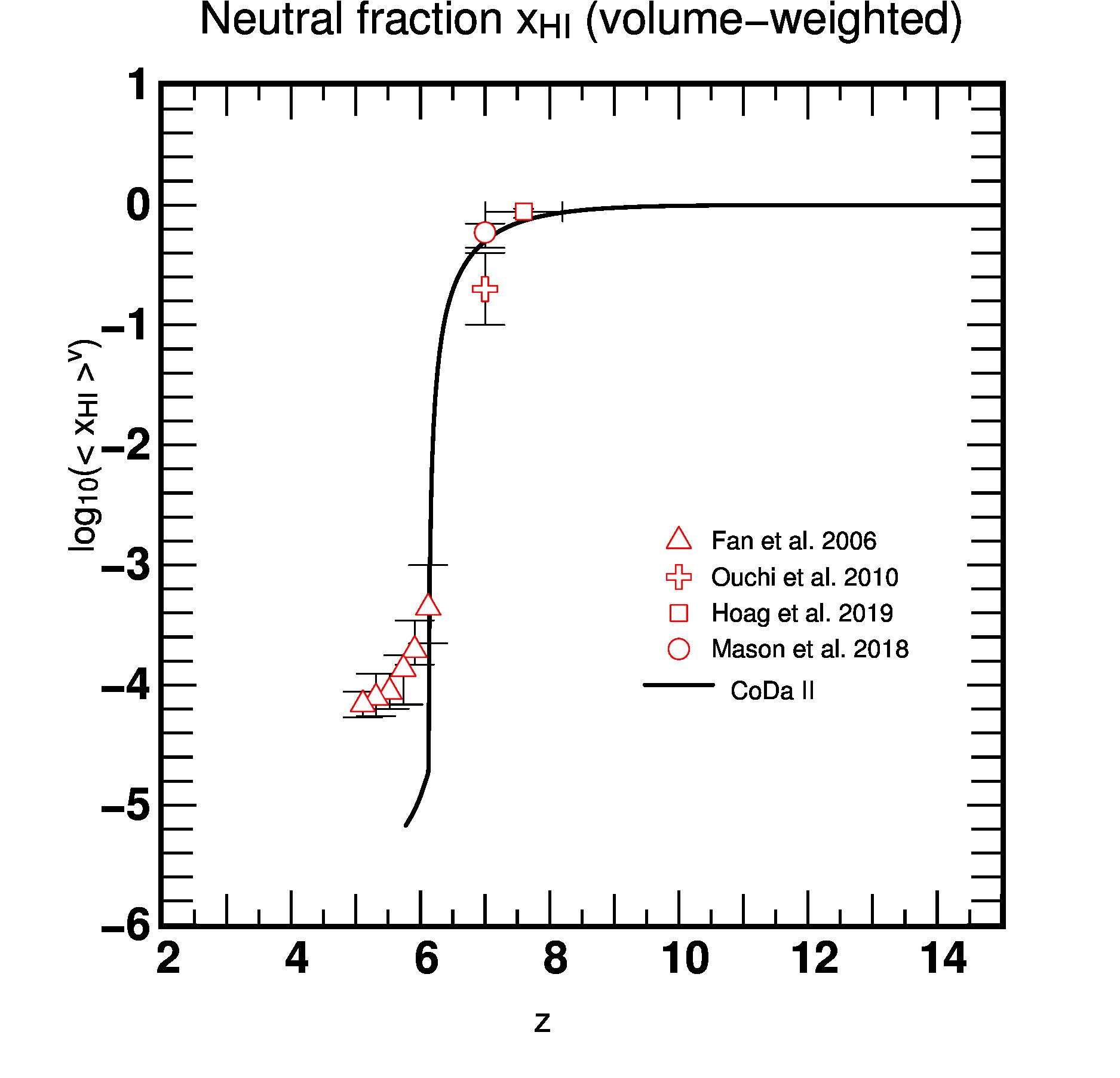}}&
  {\includegraphics[height=7.8cm,clip]{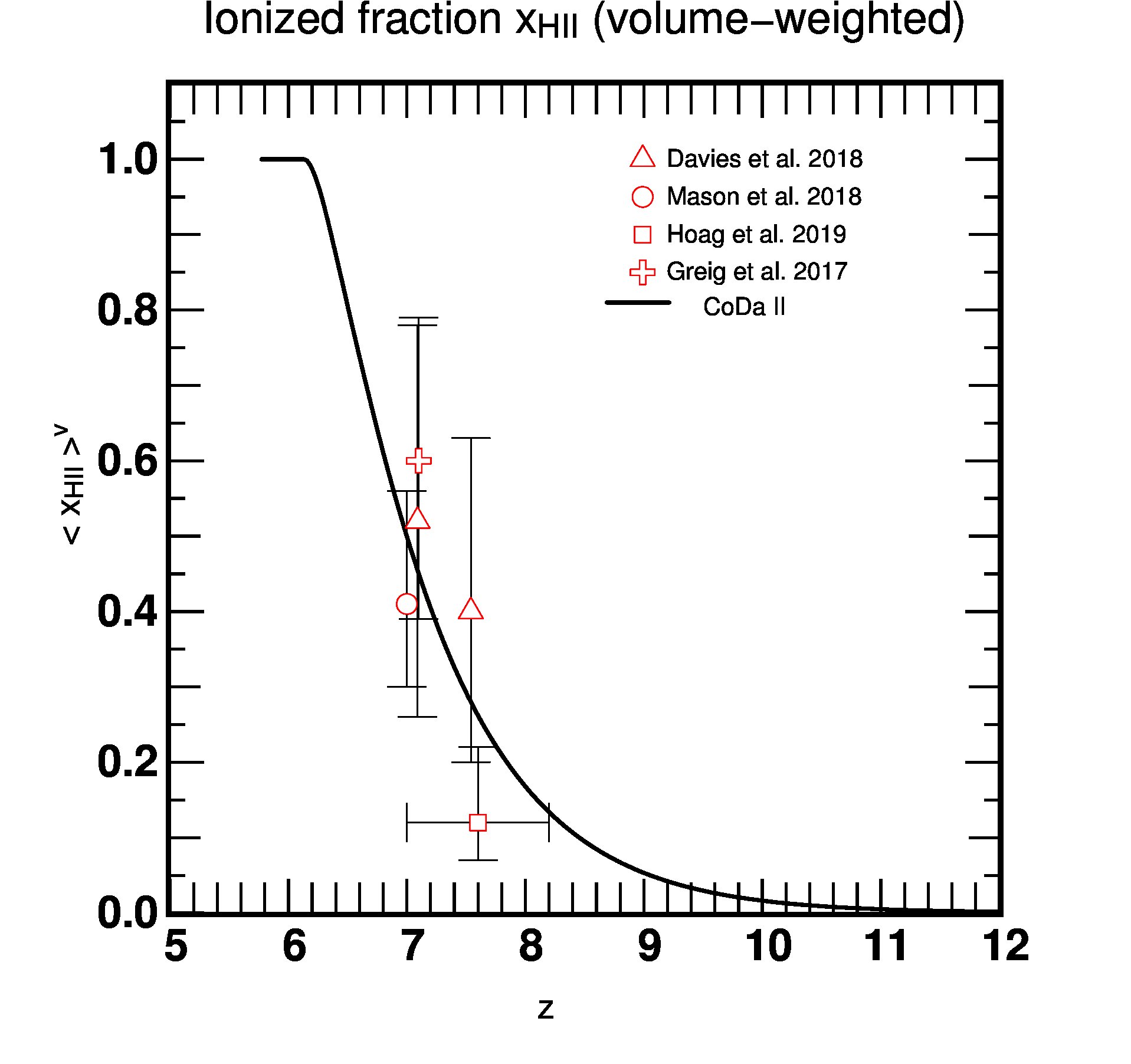}}\\
  {\includegraphics[height=7.8cm,clip]{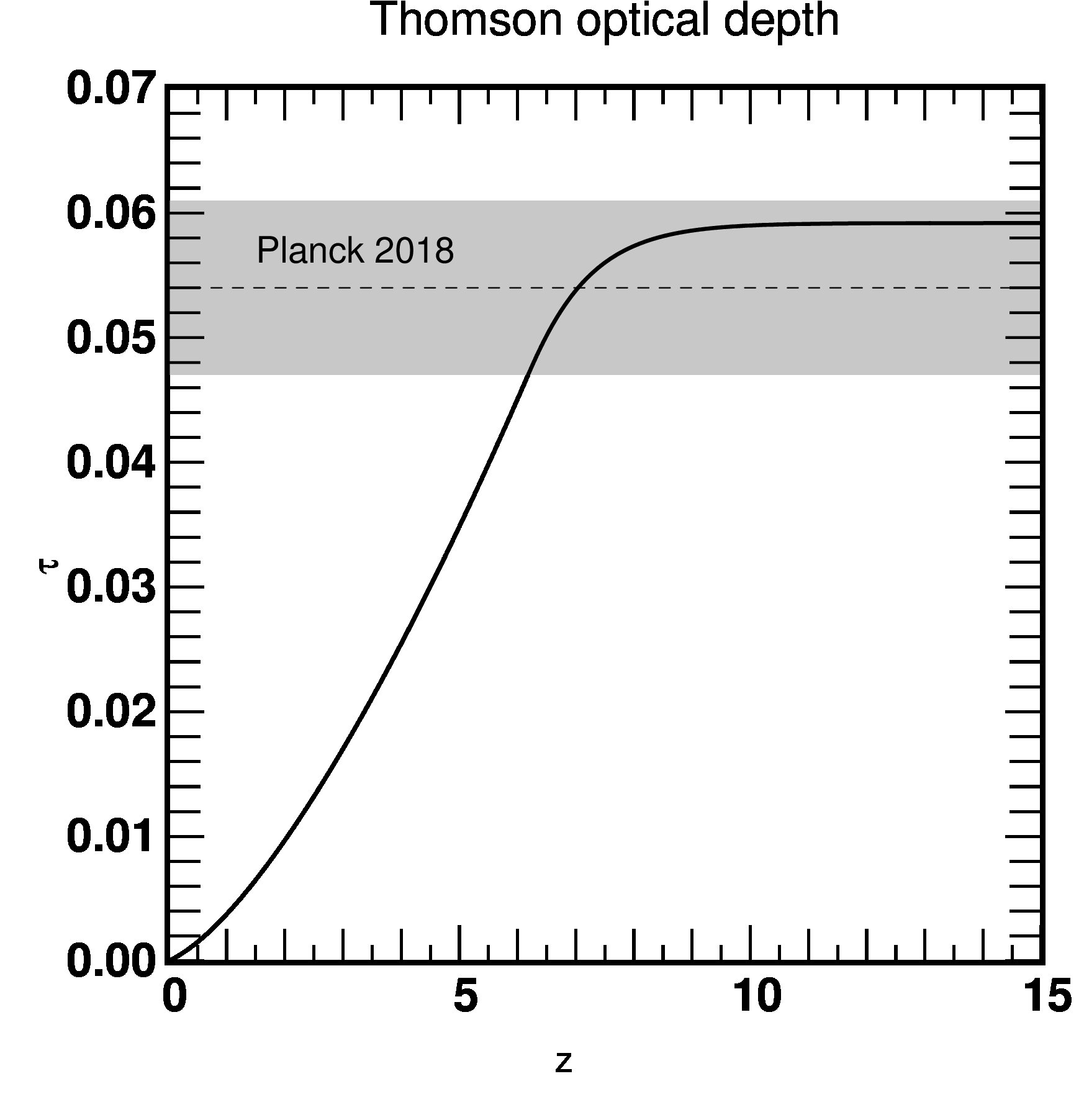}}&
  {\includegraphics[height=7.8cm,clip]{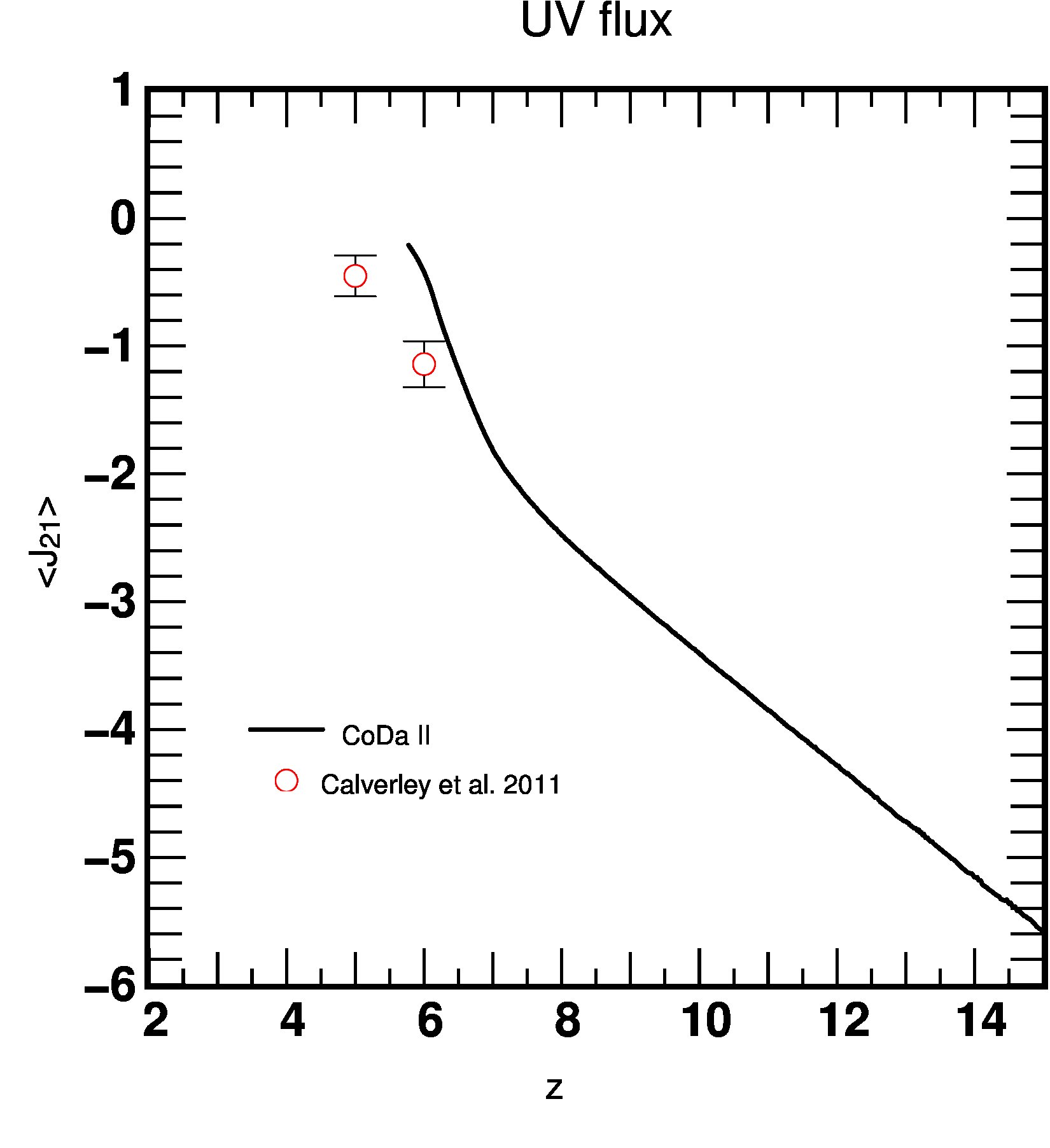}}
\end{tabular}
\end{center}
\caption{CoDa II simulation global reionization history, compared to observations: 
the evolution of the volume-weighted average of the neutral hydrogen fraction (top left) and 
ionized hydrogen fraction (top right), 
the Thomson scattering optical depth caused by free electrons in the IGM
(integrated thru the simulated universe from $z = 0$ to each $z$,
using the average ionized fraction in CoDa II at each $z$ thru the end of the EOR and
assuming the IGM was fully-ionized thereafter), compared to the value integrated
thru the entire post-recombination universe according to observations of CMB polarization anisotropy 
by Planck 2018 (bottom left), 
and the evolution of the average ionizing UV flux density (bottom right).}
\label{f:glob}
\end{figure*}

In this section we study the mean, global reionization history of the CoDa II 
simulation and compare it with available observational constraints. 
We demonstrate that the simulation is in agreement with most expectations from observations.

The most basic quantities to consider when gauging the success of a global EoR simulation 
are the evolution of the volume-weighted cosmic means of the neutral and ionized fractions, 
and the mean intensity of the ionizing radiation background. 
These are shown in Fig. \ref{f:glob}, along with several observational constraints 
from \cite{fan2006}, \cite{ouchi2010}, and the $\emph{Planck}$ CMB Thomson-scattering optical 
depth $\tau$ \citep{planck2018}.

Intergalactic H II regions grow in number and size over time 
until they overlap fully to finish reionizing the universe, before
redshift 6. As shown in the plot of mean ionized 
hydrogen fraction versus redshift in Fig. 2, the
ionized fraction was $x_{\rm HII}\sim 10^{-1}, 0.5, 0.9$, and $0.99$
at $z = 8.5, 7, 6.5$, and $6.2 $, respectively, compatible with observationally inferred values by, e.g., \cite{greig2017}, \cite{davies2018}, \cite{mason2018} and \cite{hoag2019}.
At this point, the neutral fraction plot shows a characteristic, 
very steep decrease, down to $x_{\rm HI}\sim 10^{-4.6}$, 
after which the slope becomes slightly shallower. This transition marks the end of the EoR.
According to this definition, reionization is complete in the simulation at redshift $z=6.1$, 
in agreement with the brisk drop in neutral fraction between $z=7$ and $z=6$ indicated 
by the combined measurements of \cite{ouchi2010} and \cite{fan2006}. 
The global timing of CoDa II is therefore in better agreement with these observations 
than was CoDa I. The latter ended reionization somewhat late, below redshift 5, 
although we found we could relabel the simulation redshifts of CoDa I by uniformly 
rescaling them by multiplying the simulation redshift by a factor of 1.3, 
in order to compare those results with observations or make observational predictions, 
with good agreement with these same observational constraints after this rescaling.  
Thanks to the new calibration and choice of parameters, our CoDa II simulation ended reionization earlier, by redshift 6, 
just as required by the observations, so we do not need to perform any such rescaling; 
we can compare its results here directly with observations and make direct predictions.

{The post-reionization neutral hydrogen fraction in CoDa II is about 1 dex lower than observed. 
Such an offset is not uncommon and the literature shows that simulations of the EoR can exhibit 
a variety of such departures from the observed evolution of the ionized fraction measured 
from quasar lines of sight 
\citep[e.g.][]{aubert2010,petkova2011,zawada2014,so2014,gnedin2014,bauer2015,aubert2015,codaI,codadom,ocvirk2019,Wu2019_2}.} 
This issue is likely related to the evolution of the ionizing UV background in CoDa II, shown in Fig. \ref{f:glob}. 
The surge between $z=7$ and $z=6$ correlates with the drop of the neutral fraction, 
and CoDa II overshoots the observations of \cite{calverley2011}, resulting in a lower-than-observed neutral fraction.

It is commonly accepted that resolving Lyman-limit systems is necessary to describe 
properly the population of absorbers in the IGM and predict the correct average neutral 
hydrogen fraction \citep{miralda-escude2000,shukla16}.
In particular, \cite{rahmati2017} state a physical size for Lyman-limit systems (hereafter LLSs) of $1-10$ kpc 
in their simulation. While this is compatible with the spatial resolution of CoDa II ($\sim 3.3$ physical kpc at $z=6$ 
and less at higher redshifts), LLSs may be only marginally resolved and this could have an impact on our predicted neutral 
fraction after overlap. This is sufficient to resolve the scale of Jeans-mass filtering in the photoionized IGM
after it is photoheated to $\sim 10^4$ K. However, according to \cite{emberson2013} and \cite{park16}, 
still smaller-scale structure, on the scale of the Jeans mass in the cold, neutral IGM before it is reionized,
must be resolved in order to capture fully this missing opacity of self-shielded regions, on the
minihalo scale. A suite of higher resolution simulations with RAMSES-CUDATON are required to understand 
this issue, and are beyond the scope of this paper.

\subsection{UV luminosity function}

The UV luminosity function is the other important observable simulations of the EoR must match.
To compute magnitudes, we first estimate the halo virial radius as $R_{200}$, the radius within which the average density of the dark matter is 
200 times the cosmic mean dark matter density:
\begin{equation}
    R_{200}=\left(\frac{3 \, {\rm M_{FoF}}}{4 \pi \times 200  \left< {\rho}_{\rm DM} \right> } \right)^{1/3} \, ,
\label{eq:R200}
\end{equation}
where $M_{\rm FoF}$ is the halo mass given by FoF and $\left< {\rho}_{\rm DM} \right> $ is the cosmic mean dark matter density.
We then computed the $M_{\rm{AB1600}}$ magnitudes using our BPASS Z=0.001 binary population model, assuming no dust extinction, for all the star particles associated with each individual halo, evolving the stellar populations inside each star particle over time, according to the birth-times of each particle.

\begin{figure}
  {\includegraphics[width=1.0\linewidth,clip]{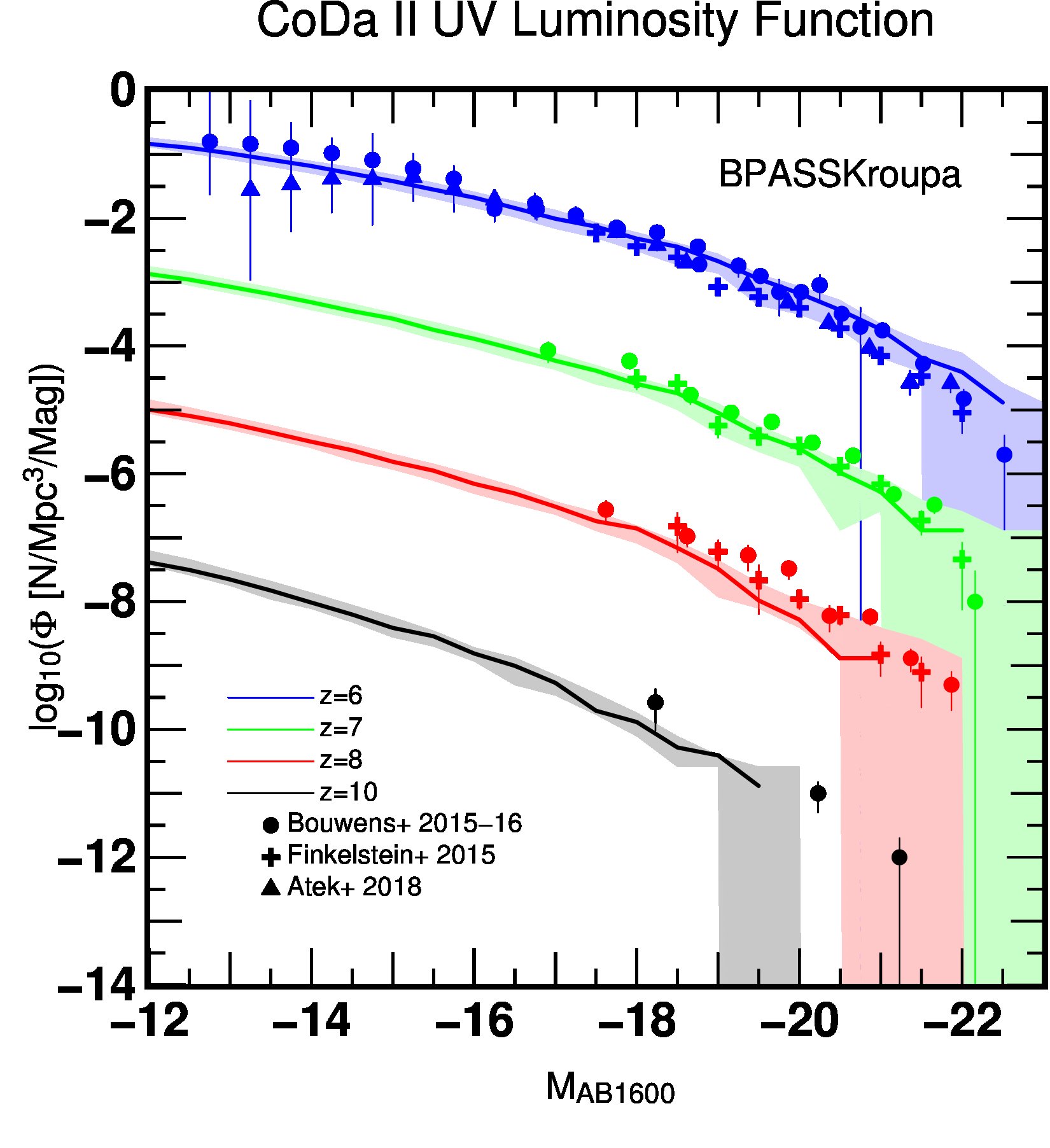}}
  \caption{{\bf CoDa II UV luminosity functions, and comparison with observations.} The full circles, crosses and triangles with error bars are the observations from \citet{bouwens2015,bouwens2017,finkelstein2015,atek2018}, respectively, at z=$[6,7,8,10]$, from top to bottom, while the shaded area and the thick line show the envelope and the median of the LFs of 5 equal, independent, rectangular sub-volumes taken in the CoDa II simulation. For clarity, the LFs have been shifted downwards by 0, 2, 4 and 6 dex for redshifts z=$[6,7,8,10]$, respectively.}
  \label{f:uvlf}
\end{figure}

The results are shown in Fig. \ref{f:uvlf}, along with observational constraints. The latter are taken from \cite{bouwens2015,bouwens2017} and \cite{finkelstein2015}, which have been shown to be in broad agreement with a number of other studies including \cite{oesch2013} and \cite{bowler2015}. We also show the $z=6$ data of \cite{atek2018}, to highlight the current uncertainties of the very faint-end obtained using gravitational lensing in cluster fields.
The luminosity functions (hereafter LF) have been shifted vertically for clarity. The shaded area shows the envelope of the LFs obtained for 5 non-overlapping, rectangular sub-volumes of the CoDa II simulation. Each of these sub-volumes spans $\sim 168,000$ cMpc$^3$ ($\Delta x=\Lbox/5$,  $\, \Delta y=\Delta z=\Lbox$, i.e. 1/5 of the full box volume), which is similar to the volume probed by CANDELS-DEEP at $z=6$. The resulting envelope therefore illustrates the expected effect of cosmic variance at $M>-20$. The thick solid line shows the median of these 5 LFs.
The CoDa II LF is in good agreement with the observations at all redshifts, although a small shift (i.e. in the sense of 
undershooting) can be seen for $z>6$, which increases at higher redshifts. 
We note, however, that given the error bars and the spread of the observed LFs, 
our results remain compatible with the observations.

\subsection{Cosmic star formation rate density}
\label{s:csfrd}
\begin{figure}
  {\includegraphics[width=1.0\linewidth,clip]{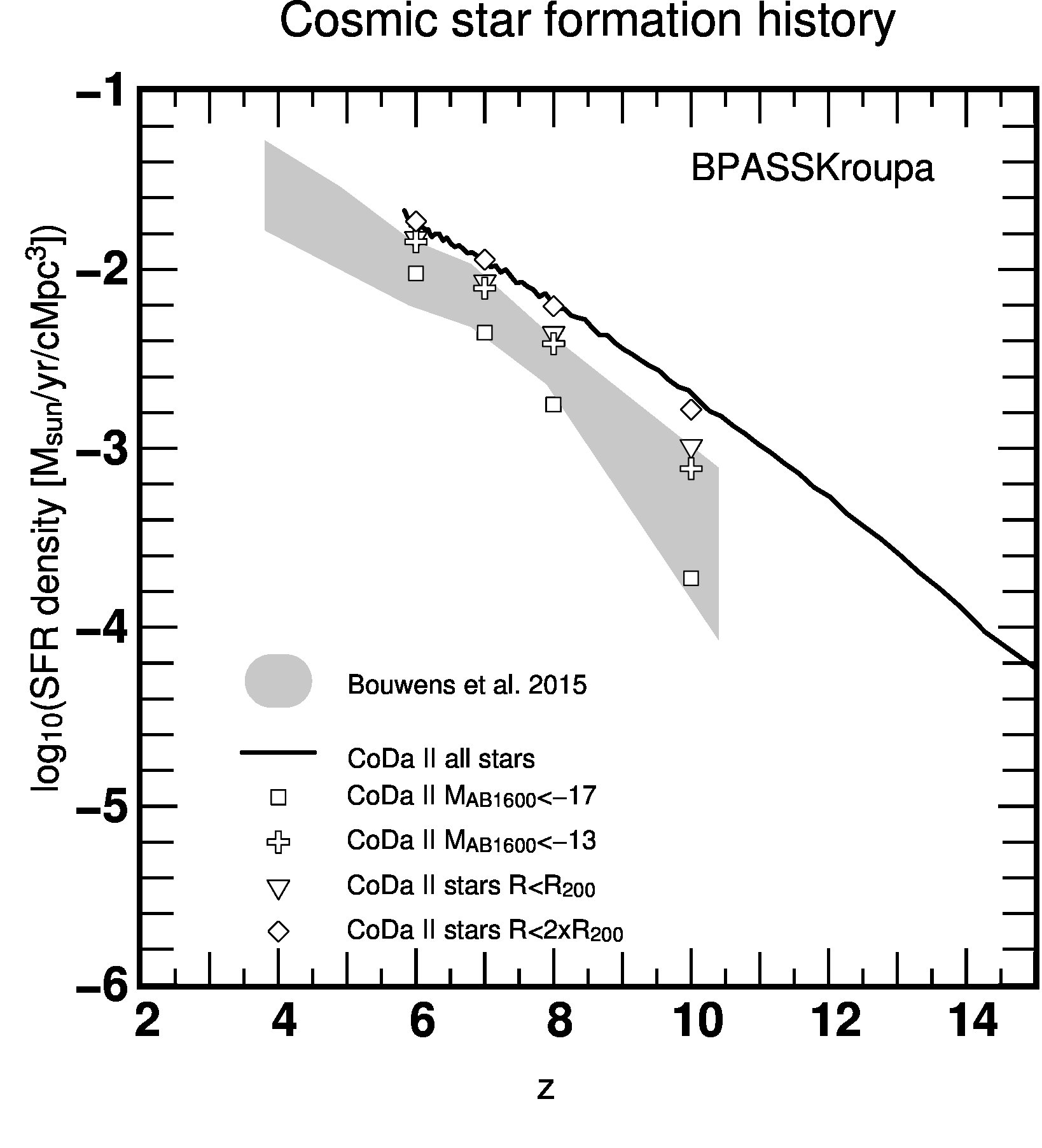}}
  \caption{{\bf Evolution of the cosmic star formation rate density (SFRD) as simulated by CoDa II, 
  compared with that inferred from observations}: (1) Using all star particles in the simulation box (thick line), (2) Accounting for haloes above realistic magnitude cuts (crosses and squares), and (3) using all star particles within the $R_{200}$ and $2  R_{200}$ spheres of haloes (triangles and diamonds). Observational result is taken from Table 7 of \citet{bouwens2015}, but with 
  the observed SFR decreased by 0.34 dex to account for the different $M_{\rm{AB1600}}$ - SFR conversion assumed here, resulting from our BPASS source model, rather than the Salpeter IMF adopted by \citet{bouwens2015}.}
\label{f:csfrd}
\end{figure}

Observations of high-redshift galaxies have been used to infer the global star formation rate
density (``SFRD") of the universe at early times, indirectly, by measuring the UV LF of galaxies and assuming
an IMF for the stars responsible for the UV starlight that escapes from each galaxy at 1600 {\AA}, 
subject to a possible correction for an unknown amount of attenuation inside each galaxy by interstellar 
dust.  In fact, there is considerable uncertainty that results from the fact that the bulk of 
the UV luminosity per galaxy is released by the most massive stars in the galaxy's IMF, 
as long as star formation continues, while the bulk of the mass in stars formed is generally assumed to
be dominated by the unseen lower-mass stars in the conventionally-assumed IMFs.  Furthermore,
once star formation ceases or slows down significantly inside a galaxy, its stellar population
will evolve passively and its UV luminosity decline, 
after its most massive stars have died a fiery death as SNe and no longer contribute their UV starlight.
In a fully-coupled radiation-hydrodynamical simulation of reionization and galaxy formation like CoDa II, 
in which the release of ionizing UV starlight and SN energy inside galaxies are very important in affecting 
the evolution both inside and between galaxies, it is only those massive stars that matter, and any inference
about the total mass of stars formed along with those massive ones is similarly subject to
an assumption about the shape of the IMF over the mass range which is well 
below the level of {10's of \Msun} which dominates the production of ionizing photons.
Nevertheless, with these caveats, we can compare our predictions for the SFRD in CoDa II
with the observational constraints by making the same underlying assumption about the IMF
for both the simulated SFR and the inferred one from the observed galaxies, as follows.

We computed CoDa II's cosmic SFRD
as the global star formation rate of the whole CoDa II volume divided by the box volume. 
The results are shown in Fig. \ref{f:csfrd}.
As a first remark, we note that the cosmic SFRD in CoDa II, as in CoDa I (O16), 
increases at all times, unlike the simulation of \cite{so2014}, 
which shows a decline at late times. In the latter, 
the authors suggested this could be due to the small box size they used. 
This may indeed be the case, since CoDa II is 
more than 95 times larger in volume.  The fact that the global SFRD 
in CoDa II increases continuously over time right through the end of the EOR was
noted previously for CoDa I in O16.  As pointed out there, this
conflicts with the prediction by \cite{barkana2000} that
the global SFR would suffer a sharp drop as the end of reionization approached,
caused by a jump in the local IGM Jeans-mass filter scale inside H II regions
as reionization overtook a significant fraction of the volume of the universe. 
While the SFR dependence on halo mass does reflect this suppression effect at lower mass,
the global SFR integrated over halo mass in CoDa II, like in CoDa I,
is apparently dominated by galaxies more massive than those whose SFR is suppressed by reionization feedback.  

In order to compare this integrated SFRD of CoDa II with the observations of high-redshift galaxies,
we must put the latter on a self-consistent footing with the assumptions of the CoDa II stellar emissivity
and IMF. As described in Section 2.2.2, our adopted value of the emissivity in ionizing photons
per stellar baryon, $ \xi_{\text{ph,}\textsc{imf}} = 1140$, assigned to each star particle 
when they form, is consistent with the ionizing emissivity of a BPASS model for
a Z = 0.001 binary population with Kroupa IMF, averaged over the first 10 Myr. 
In practice, this means that, to compare the mass in star particles formed per time in the
simulation volume with that inferred from observations of the UV LF of high-redshift galaxies,
those observations must be interpreted in terms of the same assumed IMF when
inferring the observed SFRD and its cosmic evolution. 
Also shown in Fig. \ref{f:csfrd} are the observational constraints taken from  \cite{bouwens2015}: 
the grey area shows the envelope  bracketing the dust-corrected and dust-uncorrected SFRD's. 
We decreased this observed SFRD by 0.34 dex to account for the different $M_{\rm{AB1600}}$ - SFR conversions 
resulting from our assumption of the BPASS model, rather than the Salpeter IMF for single-stars 
assumed by \cite{bouwens2015}. We computed this offset by applying the methodology of 
\cite{madau1998} to our BPASS source model. This involves computing the 1600 {\AA} luminosities of 
a population undergoing an exponentially decaying burst of star formation, 
resulting in a conversion factor:
\begin{equation}
L_{1600}= \, {\rm const} \, \frac{{\rm SFR }}{\Msun {\rm yr^{-1}}} \, {\rm ergs \, s^{-1} \, Hz^{-1}} \, ,
\end{equation}
where const $=1.75 \times 10^{28}$ with our source model, and const $=8 \times 10^{27}$ in B15.

  
The CoDa II total SFRD (thick solid line) overshoots the observations by about 0.1 dex at $z=6$. 
However, the cosmic SFRD inferred by B15 only includes galaxies with ${M_{AB1600}}<-17.7$, 
and does not account for fainter (undetected) objects. Therefore, to compare our CoDa II results
to B15 in a fair way, we need to apply similar magnitude cuts to both. With such cuts 
(squares in Fig. \ref{f:csfrd}), CoDa II falls within the observationally-favored SFRD at $z=6$, 
and very close to the observations at higher redshifts.

At $z=6$, the SFRD inferred from the CoDa II sample with ${M_{\rm{AB1600}}}<-17$  
is 2.5 times smaller than the total SFRD in the box, and the offset grows with increasing redshift.
However, not all that offset is due to the magnitude cuts:  the triangles in Fig. \ref{f:csfrd} show 
the cosmic SFRD obtained when considering only those star-particles which are located within
a sphere of radius $R_{200}$ centered on the center of each halo, as opposed to considering all the 
star particles in the box. Even when considering all halos, the $R_{200}$-bounded SFRD is always smaller 
than the full-box SFRD, showing that some of the star particles are located either beyond 
$R_{200}$ for haloes detected by the FoF halo-finder or else are not associated with haloes
because their haloes are below our assumed lower halo mass limit of $10^8$ \Msun (246 dark matter particles). { Haloes, in particular at the high redshifts studied here, are strongly triaxial, 
so that their particles and stars may extend beyond a sphere of radius $R_{200}$. This aspect is illustrated in more detail in Appendix \ref{s:orphanstars}, where we show that stars beyond $R_{200}$ are still associated with the most nearby dark matter halo. We further confirm this by considering all stars within $2 R_{200}$ (diamonds in Fig. \ref{f:csfrd}), yielding a SFRD which matches fhe total full box SFRD at redshifts below z=8. Only at z=10, some of the stars may be associated to small haloes below the assumed limit.}

Finally, comparing the ``detectable'' CoDa II galaxy population with the total SFRD within the
$R_{200}$ spheres of all haloes, we find that the CoDa II sample with ${M_{\rm{AB1600}}}<-17$ at $z=6$
accounts for about 63\% of the halo $R<R_{200}$ SFRD, and only 18\% at $z=10$. 
On the other hand, in the case of deeper detection limits such as in lensing cluster fields 
where lensing magnification makes it possible to detect fainter haloes with
${M_{\rm{AB1600}}}<-13$, these fractions increase to 94\% and 76\%, respectively.

\subsection{The very faint end of the luminosity function}

\begin{figure}
  {\includegraphics[width=1.\linewidth,clip]{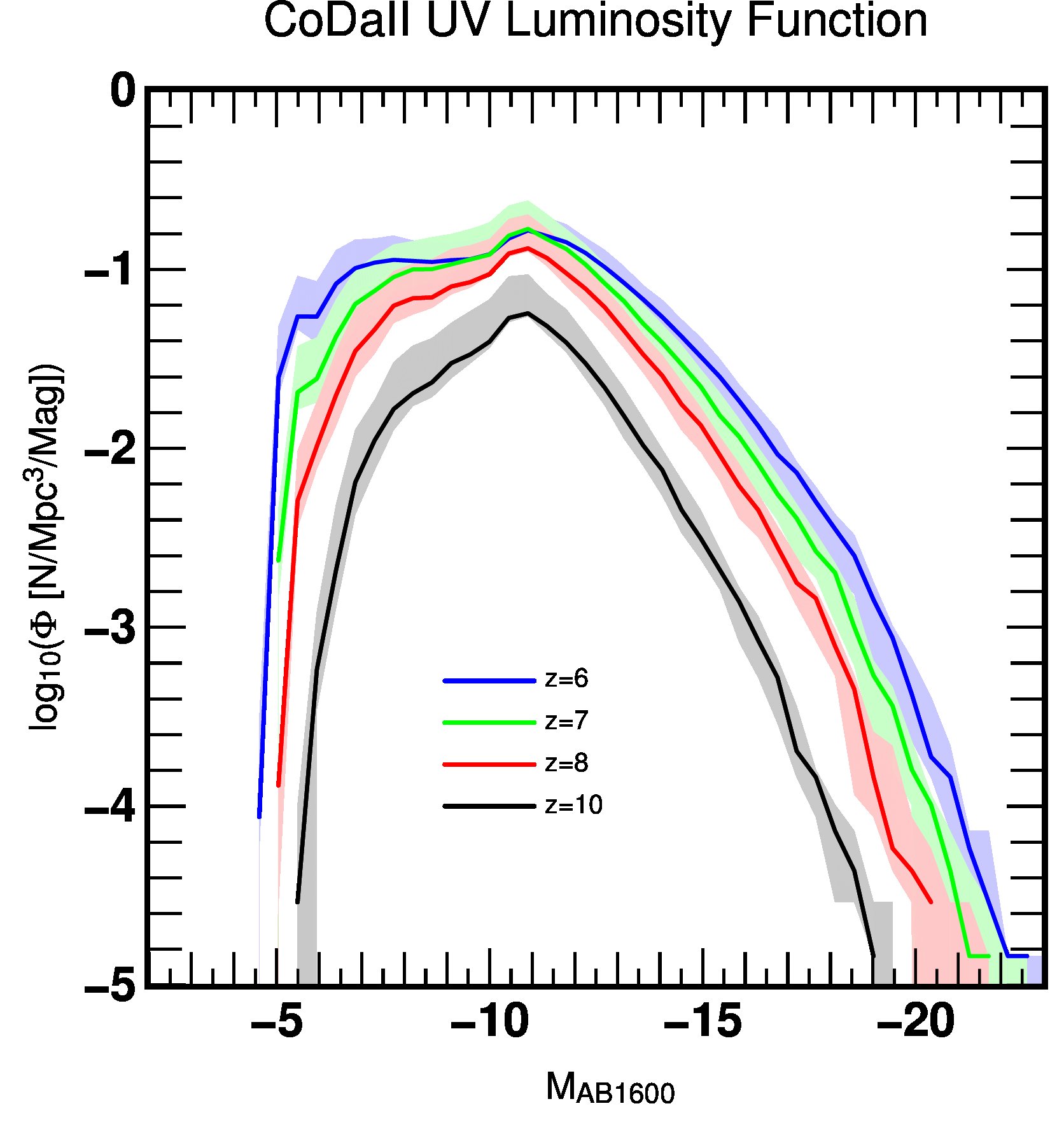}}
  \caption{CoDa II UV luminosity functions, including the very faint end, at 4 epochs. The shaded area and the thick line show the envelope and the median of the LFs of 5 equal, independent, rectangular sub-volumes taken in the CoDa II simulation. The full LFs are given in a table in appendix.}
  \label{f:gluvlf}
\end{figure}

We can also use CoDa II to investigate the LF at magnitudes fainter than currently observed. We show in Fig. \ref{f:gluvlf} the whole LF, going down to $M_{\rm{AB1600}}=-5$, for 4 redshifts. The progressive build-up of the LF with cosmic time is clearly apparent, both at the bright end, with more and more massive haloes hosting vigorous star formation, and at the faint end, made of passively evolving galaxies.
The maximum of the LF is reached at $M_{\rm{AB1600}}\sim -11$ at all redshifts. At redshifts lower than 10, a plateau develops faintwards of -11, extending to fainter magnitudes as time goes by. At $z=6$, the plateau extends down to $M_{\rm{AB1600}}=-6$. We show in Fig. \ref{f:massmag} that this [-11,-6] magnitude range corresponds to masses of $10^8-10^9$ \Msun, i.e. the mass range where SFR suppression by ionizing radiation takes place (see Sec. \ref{s:SFR}). 
Indeed, if haloes below $\sim 10^9$ \Msun have their SFR's suppressed after they
are overtaken by reionization and, thereafter, have only the luminosity associated with the passive evolution of their
low-mass stars (after the high-mass stars finished their lives as SNe), we expect the number density of galaxies
with magnitudes fainter than $M_{\rm{AB1600}}=-11$ to grow continually, as more and more volume is overtaken by reionization, as shown by our figure.

The full LFs for these 4 redshifts are given in the appendix in tabular form.

\subsubsection{Halo Mass and UV luminosity}
\label{s:hmuv}

\begin{figure*}
  {\includegraphics[width=0.49\linewidth,clip]{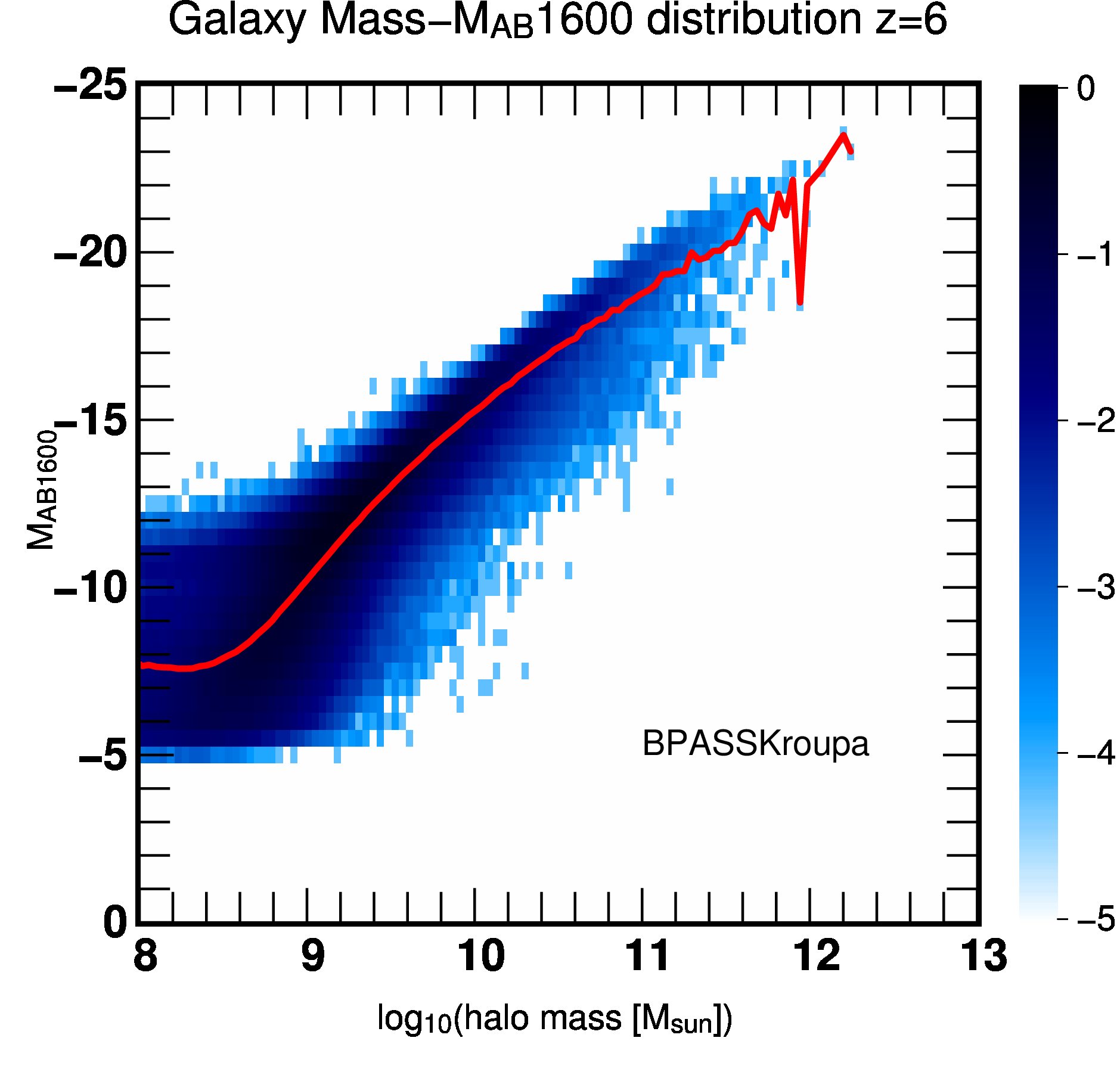}}
  {\includegraphics[width=0.49\linewidth,clip]{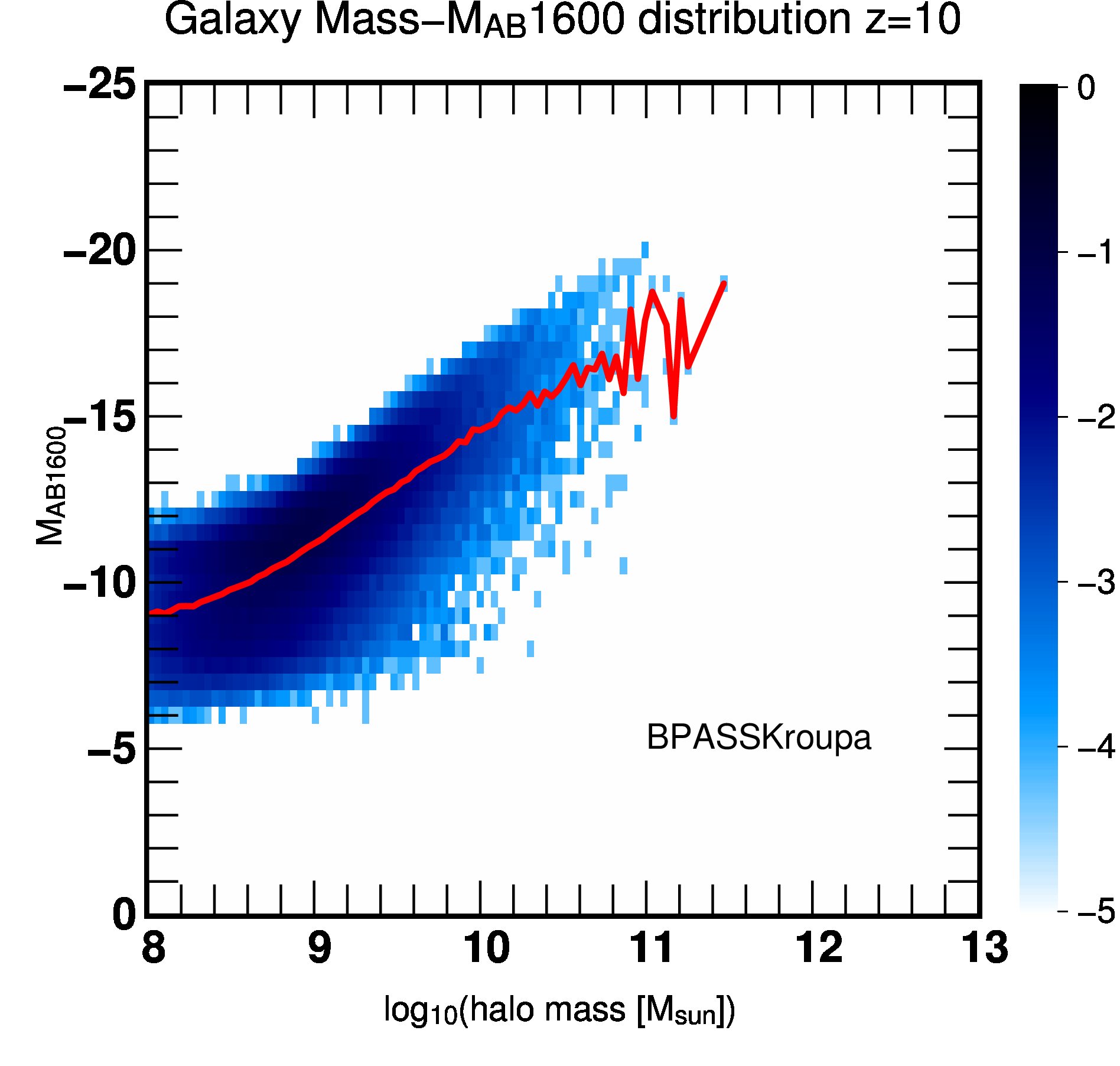}}
\caption{CoDa II galaxy mass - magnitude distributions at z=6 and z=10. The color indicates the galaxy comoving number density in N/Mpc$^3$/Mag/log(M$_{\odot}$). The red line indicates the average Magnitude for each mass bin.}
\label{f:massmag}
\end{figure*}

The results of CoDa II show that the UV continuum luminosity of a galaxy is correlated with the mass of its galactic halo. 
The average luminosity of halos of the same mass at a given redshift increases with halo mass,
consistent with the correlation found for Coda I (O16).  
This is shown for CoDa II in the panels of Fig. \ref{f:massmag}, which correspond to the galaxy populations 
of the $z=6$ and $z=10$ UV LF's in Fig. \ref{f:uvlf}. Moreover, for each halo of a given mass, the luminosity can fluctuate 
significantly over time, which is reflected in the vertical spread of the distributions in Fig.\ref{f:massmag} . 
At z=6, the vertical dispersion increases with decreasing halo mass.
Moreover, the spread at low mass is larger at $z=6$ than $z=10$. The increase in dispersion 
is not driven by the brightest galaxies of the low-mass haloes. Indeed, the brightest galaxies residing in a 
haloes of mass a few times $10^8$ \Msun have ${M_{\rm{AB1600}}} \sim -12$, in both snapshots, 
and contain a couple of 1-2 Myr old elementary stellar particles in CoDa II.
By contrast, the faintest galaxies of the $z=10$ snapshot have ${M_{\rm{AB1600}}} \sim -5.5$, 
while the faintest galaxies of the $z=6$ snapshot are as faint as ${M_{\rm{AB1600}}} \sim -5$. 
This is the signature of passively evolving galaxies: their star formation has stopped, 
and their stellar populations progressively fade, making them fainter as redshift goes down.

\subsection{Impact of radiative feedback on galaxy formation}
\label{s:SFR}
\begin{figure}
\begin{center}
 {\includegraphics[width=1.\linewidth,clip]{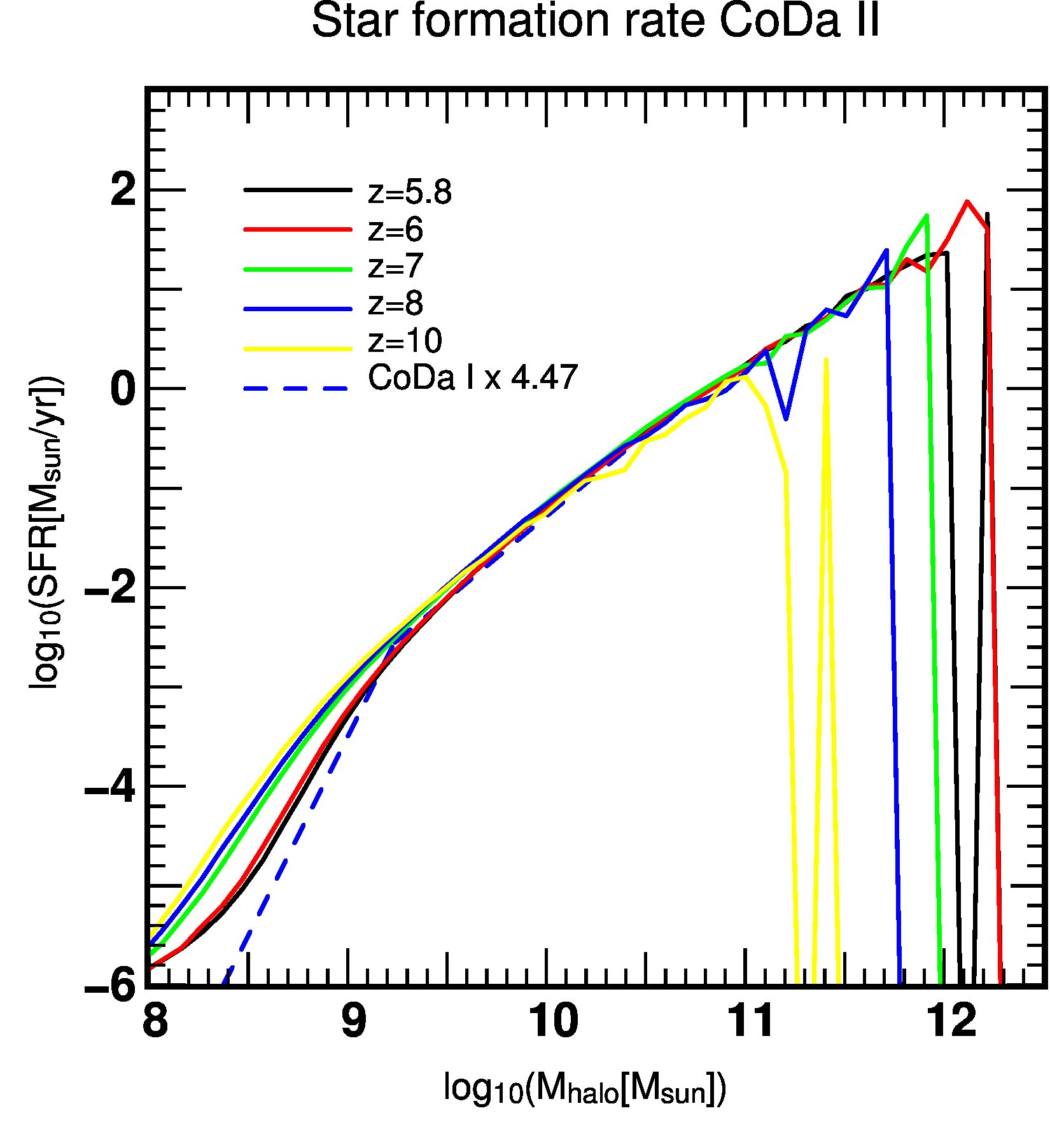}}  
\end{center}
\caption
{Instantaneous star formation rate per halo as a function of instantaneous halo mass, for various redshifts. The instantaneous SFR is computed as the stellar mass formed within an $R_{200}$ radius sphere centered on the dark matter halo center of mass, during the last 10 Myr, divided by a duration of 10 Myr. The solid lines show CoDa II data while the dashed line shows the post-overlap CoDa I data, multiplied by a factor to make it match the high mass regime end, for easier comparison.}
\label{f:SFR}
\end{figure}

\begin{figure}
  {\includegraphics[width=1.075\linewidth,clip]{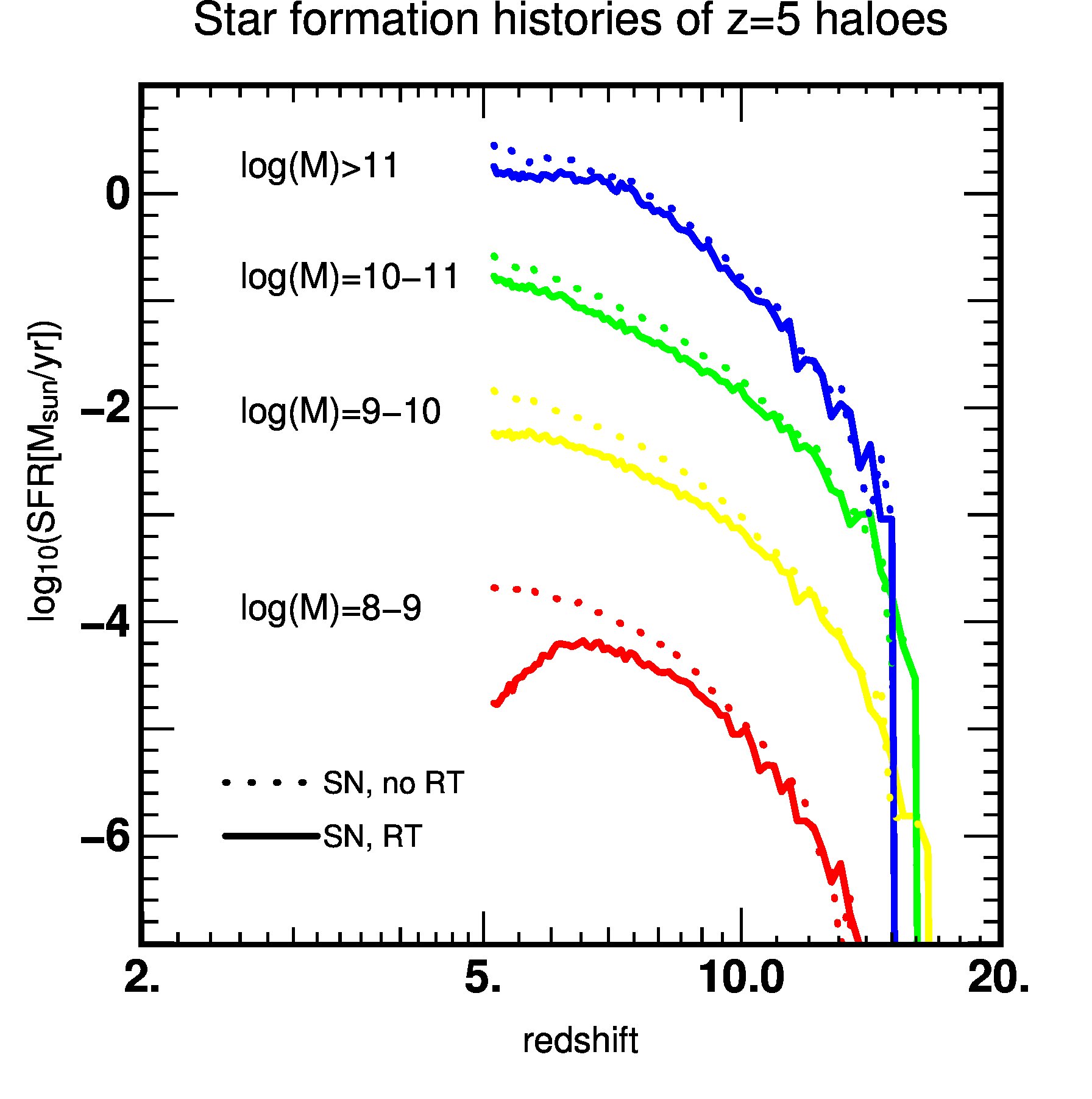}}
  \caption{
    Star formation histories of 4 halo mass bins for two simulations in a test box 8 \chmpc on a side. {\em Solid line}: with full fiducial physics (supernovae and radiative transfer), {\em dotted line}: with supernovae only, no radiation.  The mass bins correspond to the haloes final mass (i.e. halo mass measured in the final simulation timestep, at z $\sim5$.)}
  \label{f:fbstudy}
\end{figure}

As described in Sec. 1, the CoDa suite of fully-coupled radiation-hydrodynamics simulations
was designed to address the long-standing question of how galaxies that released the 
UV H-ionizing radiation that reionized the universe were affected by the
photoheating that accompanied photoionization during the EoR, acting back on their gas 
content and star formation rates, and on the progress of reionization, itself. In O16, 
we summarized the globally-averaged effect of this feedback on the star formation rates 
of halos of different mass at different redshifts in CoDa I, showing that the SFR in 
low-mass haloes below a few times $10^9 \text{ M}_\odot$ was reduced toward the end of reionization, 
as more and more of the volume was reionized. 
In \cite{dawoodbhoy2018}, we studied these CoDa I results in more detail, 
going beyond the globally-averaged results to identify
the causal connection between the suppression of star formation in low-mass haloes and 
reionization by showing that SFR suppression
followed the local time of arrival of reionization at the location of each galaxy.
In that case, we showed, reionization was not
limited to the late stage of reionization, as it appeared from the globally-averaged results in O16, 
but was happening throughout the EoR, at different times in different places. 

In \cite{dawoodbhoy2018}, we summarized the literature which had attempted to characterize 
this feedback effect and establish the mass range of halos subject to suppression, 
by variety of approximations over the years, including simulations. We refer the reader to that summary and 
the references therein, for this background. In the context of 3-D galaxy formation simulations, 
a number of authors have attempted to address this impact of radiative feedback on galaxy formation, 
by adopting a pre-determined, uniform UV background like that modelled by \cite{haardtmadau96,haardt2012}), 
and more recently with fully-coupled RHD simulations \citep{pawlik2013,wise2014,jeon2014,rosdahl2015,aubert2015,pawlik2015,codaI,rosdahl2018,dawoodbhoy2018,Wu2019_2,katz2019}, 
investigating a possible suppression or reduction of star formation in low-mass galaxies.  
Here we will highlight some of the
results of the CoDa II simulation on this issue and compare them with those in CoDa I.  

We computed the instantaneous SFR of CoDa II haloes as the stellar mass formed within a 
sphere of radius $R_{200}$ centered on the dark matter halo center of mass, during the 
last 10 Myr, divided by a duration of 10 Myr. Fig. \ref{f:SFR} shows the instantaneous SFR that 
results, as a function of the instantaneous mass of the dark matter halo, for several redshifts.

There is a general trend for more massive haloes at all epochs to form more stars, as seen in 
\cite{ocvirk2008} and O16. Here we find a trend: SFR $\propto M^{\alpha}$ for $M>10^{10}$ \Msun   
with a slope $\alpha \sim 5/3$, just as in CoDa I. This slope is compatible with the values 
found in the literature. i.e $1<\alpha<2.5$, in numerical and semi-analytical studies, such as \cite{tescari2009,hasegawa2013,yang2013,gong2014,aubert2015,codaI}. 
Moreover, Fig. \ref{f:SFR} shows a smooth decrease in SFR for the low mass haloes, 
around $\sim 2 \times 10^{9}$ \Msun, relative to this slope at higher mass. Below this mass, 
the SFR-mass relation is slightly steeper, and gets even steeper at lower redshifts, as reionization advances. 
This steepening at late times during the EoR is a signature of suppression by reionization feedback, 
also seen in the globally-averaged results from CoDa I. However the strength of the suppression is stronger in CoDa I than in CoDa II.

In order to understand the origin of this evolution in the SFR-mass relation, 
in particular with respect to the relative importance of supernova feedback and 
radiative feedback, we performed two additional control simulations, with the same 
parameters as CoDa II, but in a smaller box, 8 \chmpc on a side, isolating the feedback physics:
\begin{itemize}
\item{{\em SN, RT}: this simulation has the exact same physics as CoDa II, i.e. SNe and radiation.}
\item{{\em SN, no RT}: this simulation was run with SNe but without radiation 
(and therefore does not undergo global reionization).}
\end{itemize}
Both simulations ran to redshift $z=5$. Fig. \ref{f:fbstudy} shows the star formation history of 
haloes present in the final snapshot at $z=5$, binned according to their final masses at this redshift. 
Again, the trend of increasing SFR with halo mass is clear. Also, expectedly, since $z=5$ haloes were 
less massive at, say, $z=15$, their SFR is lower at higher redshift. For all mass bins, the SFR of the 
run with radiation is smaller than without radiation. While the effect is very small in the highest mass bin, 
it gets stronger the lower the mass. The lowest mass bin, in particular, shows a striking behaviour: 
the SFR increases with time, and then starts to decrease around $z=6.5$, i.e. during overlap and shortly 
before the end of reionization. This is markedly different from the run without radiation, in which the 
low-mass haloes have a monotonically increasing SFR. This demonstrates that the UV ionizing radiation 
suffusing the Universe during the EoR is indeed the cause of the withering SFR of the low mass haloes in CoDa II.

This gradual suppression reflects a radiation-driven reduction in the baryonic fraction of low-mass 
haloes \citep{dawoodbhoy2018,sullivan2018}. In contrast, more massive haloes are able to retain their 
gas and keep on forming stars at a very similar rate pre- and post-overlap. We note that the reduction 
in the SFR of low-mass haloes due to global photo-heating during reionization is less severe than in CoDa I. 
Indeed, the sub-grid model for star formation in CoDa I included an additional criterion, allowing stars 
to form only in cells colder than $T=20 000$ K, thereby making a smaller amount of cells in the simulation 
eligible for star formation after local reionization occurs. This more stringent requirement led to a more 
severe suppression of star formation in low-mass haloes (dashed curve in Fig. \ref{f:SFR}). This requirement 
was relaxed in CoDa II, where star formation is purely based on a gas overdensity threshold and an efficiency parameter 
(see Table \ref{t:sum}). It is interesting to note that suppression nevertheless still clearly happens
in Coda II at low masses, although in a slightly less severe form.

We performed a resolution study, presented in Appendix \ref{s:resstudy}, where we show that increasing the spatial resolution of our setup by up to a factor 4 in space and 64 in mass (although in a smaller test box), results in similar star formation suppression in the low mass halo range ($10^8-10^9$ \Msun) after reionization. A quick comparison with the literature, e.g. \cite{pawlik2015} and \cite{Wu2019_2}, shows reasonable agreement. Therefore, star formation suppression in CoDa II is not likely to be an artefact due to numerical resolution.


\subsection{Environment and the timing of SFR suppression in low-mass galaxies}
\label{s:SFRdel}
\begin{figure*}
\begin{center}
  {\includegraphics[width=0.49\linewidth,clip]{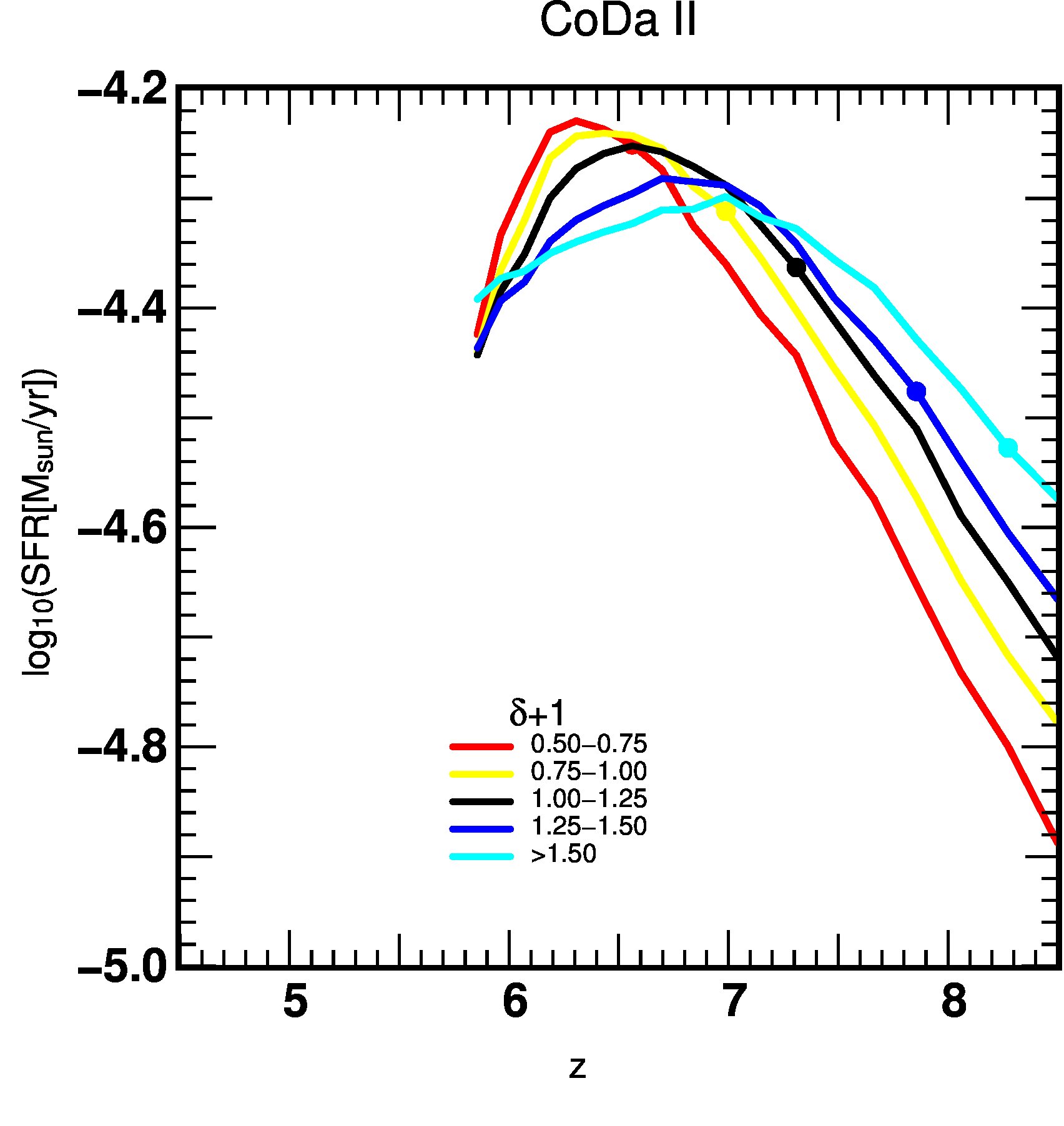}}
  {\includegraphics[width=0.49\linewidth,clip]{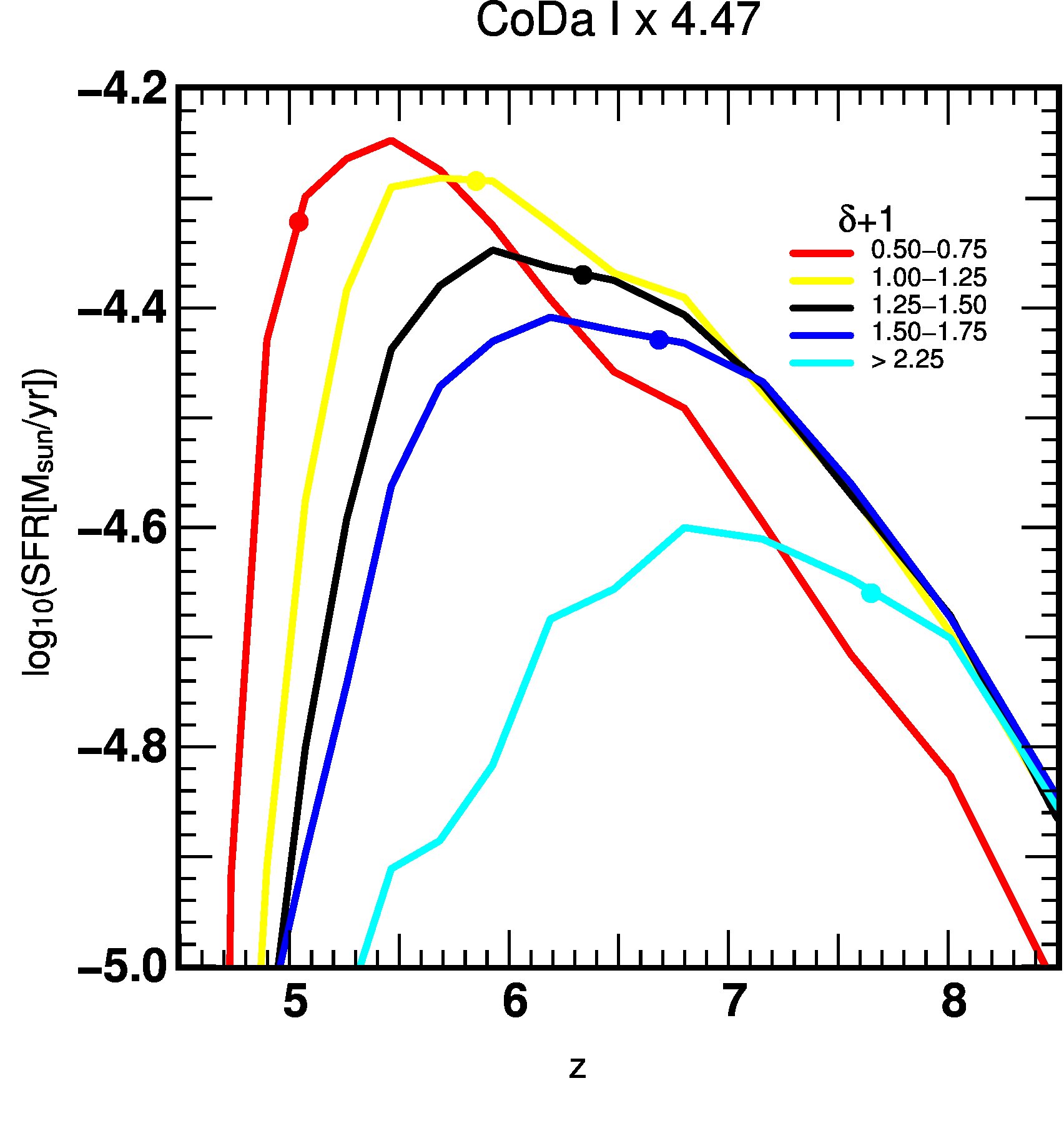}}

\end{center}
\caption{Avergage star formation histories of low-mass haloes ($2.5\times 10^8$\Msun $<M<7.5\times 10^8 $\Msun) in CoDa II (left) and CoDa I (right), for halos present in that mass bin just after global reionization ended (z=5.8 for CoDa II and z=4.44 for CoDa I) for different environments, 
 quantified by their overdensity $\delta +1 = \rho / \bar{\rho}$ computed on 4 \chmpc sub-boxes. 
  The filled circle on each curve shows the average reionization redshift (when ionized fraction
  was 0.5) of each overdensity bin. Note that the overdensity bins we used are slightly different between the two simulations, because the lower reionization redshift of CoDa I results in a higher maximum overdensity.}

\label{f:SFRdel}
\end{figure*}

Reionization is a patchy process, as shows Fig. \ref{f:bigmap}. Therefore in the context of the radiative feedback demonstrated in Sec. \ref{s:SFR}, we may expect different environments (galaxy cluster or group progenitor, galaxy neighbourhood, voids), to affect low mass galaxies' star formation rates with different strengths and at different times. The CoDa II simulation is well suited to such investigations thanks to the variety of environments it contains and its self-consistent modelling of the inhomogeneous progress of reionization.
Here we wish to illustrate this aspect, while leaving a more complete, detailed study in the spirit of \cite{dawoodbhoy2018} for a later paper.

We computed the overdensity in which CoDa II galaxies reside at the end of the simulation, as $\delta + 1 = \rho / \bar{\rho}$ averaged in sub-boxes of 4\chmpc on a side. This is the typical scale of a galaxy group progenitor, and therefore this quantifies the environment at an intermediate, meta-galactic scale.
We divide our galaxy population into 5 overdensity bins 
and compute the average star formation histories of CoDa II low-mass 
haloes ($2.5\times 10^8$\Msun $<M<7.5\times 10^8 $\Msun at $z=5.8$. 
The results for CoDa II are shown in the left panel of Fig. \ref{f:SFRdel} 
for the 5 overdensity bins. Each curve shows the same characteristic parabolic 
shape seen for low-mass haloes in Fig. \ref{f:fbstudy}, with a timing and amplitude clearly 
correlated with the local overdensity. 

In particular, haloes in dense environments have a more intense early (z=8) star formation history. This is likely caused by the inherently faster formation of their dark matter haloes and therefore larger mass accretion rates, as compared to underdense regions, as shown in \cite{maulbetsch2007}.

The filled circle overplotted on each curve shows 
the average redshift at which the cells in that overdensity bin reached the point of
ionized fraction 0.5. 

In the highest density environments, 
the star formation in low-mass haloes starts early, culminates around $z=7$, and then decreases.

In underdense regions, the same scenario unfolds, with a delay: star formation starts later, 
reionization happens later, and therefore the average star formation rate peaks and 
decreases later as well. 

Strikingly, at $z\sim 6.2$, low-mass haloes in underdense regions form 
stars at a higher rate than their counterpart in overdense regions. However, this episode is short-lived, 
and as the epoch of reionization comes to an end, the average star formation rates of low-mass 
haloes seem to converge to a similar value independent of their environment.

These results for CoDa II are consistent with the results found for CoDa I 
by \cite{dawoodbhoy2018}, both for the correlation of the local reionization times (or redshifts)
with local overdensity and the suppression of SFR in low-mass haloes following their local
reionization time.  We note that in \cite{dawoodbhoy2018}, the definition used for reionization time
(or redshift) was the $90\%$-ionized time, which is somewhat lower redshift than the $50\%$-ionized 
time used here.  Hence, the delay found here for CoDa II between the half-ionized redshift and
the redshift at which the SFR peaks and then declines for a given overdensity bin, due to
suppression, is larger here due to this choice of half-ionized reionization redshift. 

To make a direct comparison here between CoDa II and CoDa I,
we show in the right panel of Fig. \ref{f:SFRdel} the same analysis, 
performed this time on CoDa I data, for the same range of halo masses 
($2.5 \times 10^8$\Msun $<M<7.5\times 10^8 $\Msun) just after the end of global reionization, i.e. $z=4.4$. 
The SFRs for CoDa I have been multiplied by 4.47 as in Fig. \ref{f:SFR} to make them comparable in amplitude to CoDa II results. 
Because of the lower redshift, the maximum overdensity of the simulation is higher. The behavior of low-mass 
galaxies share several similarities between the two simulations: in the mass range considered, 
the SFRs of all overdensity bins rise, culminate and decay, correlated with the average
reionization times of their bin. Also, the earlier suppression taking 
place in overdense regions is confirmed in CoDa II, although in a less severe form, 
as was already noted in Sec. \ref{s:SFR}. We note however that we are unable to predict what 
would happen if CoDa II had run to lower redshift. It could be that at $z=5$, CoDa II low-mass haloes 
would end up with very low SFRs, comparable to those of CoDa I. 
Further simulations will be needed to address this aspect conclusively.

\subsection{Mapping the inhomogeneous timing of reionization in the Local Universe: "islands in the stream".}

\begin{figure*}
    \centering
    \includegraphics[width=\textwidth]{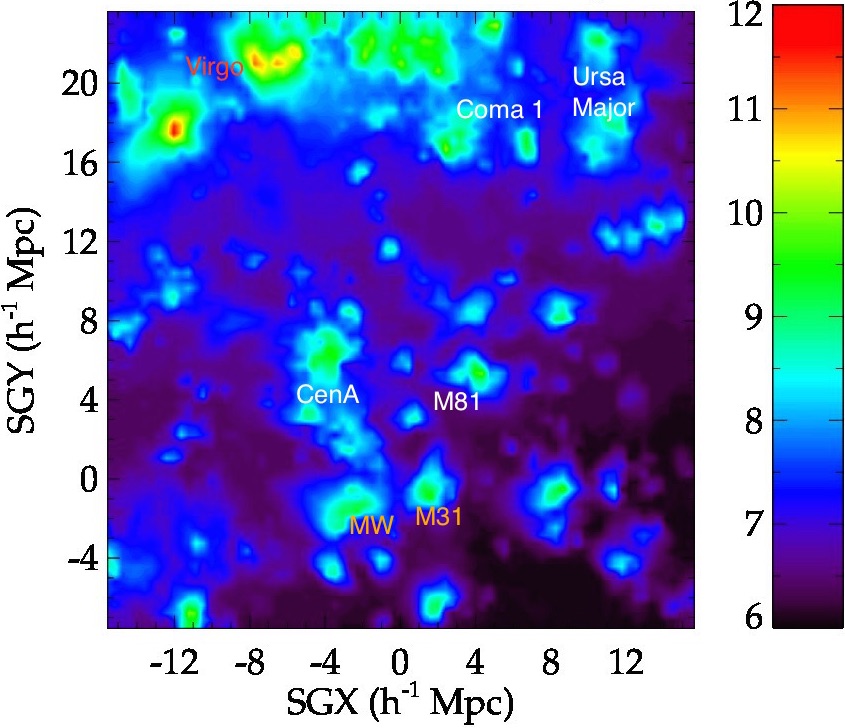}
    \caption{{\bf Reionization map of the Local Universe: Islands in the Stream}. The redshift at which each
    of the $1024^3$ grid cells first reached an ionization of 50\% is displayed in a planar-slice
    1 $h^{-1}$ cMpc thick through the CoDa II simulation volume, labelled with the names of progenitors
    of familiar objects in the Local Universe today at their respective locations in the map.}
    \label{fig:islands}
\end{figure*}

The effects of reionization arrive at different times in different places, not only globally, but
within the Local Universe, as well.  The CoDa II simulation, with its constrained initial
conditions, was designed to capture this aspect of the reionization of the Local Universe.
To illustrate it, we computed the reionization redshift of CoDa II gas cells as the redshift at which the ionized fraction of the cells reaches 50\% for the first time. This was done using the level 10 ionized fraction of the simulation, yielding a reionization redshift field of 1024$^3$ cells (4096$^3$ is level 12). Using the level 10 instead of level 12 reduces the memory requirements of this process by 64, but does not impact our results. A reionization map is then created by  
a slice through this reionization redshift field
1 $h^{-1}$cMpc thick, shown in Fig. \ref{fig:islands}.
The progenitors of several major Local Universe objects analogs were traced back from z=0 to z=6 using the CoDa II-DM2048 simulation. They are labelled on Fig. \ref{fig:islands} to allow us to gauge their reionization history and radiative influence region during the epoch of reionization in the Local Universe.


The map shows local maxima of $\zreion$, 
which are ``islands" of early reionization surrounded by a ``sea" in which reionization happened
at a later time. Such regions are internally reionized, in an inside-out pattern.


As can be seen in Figure \ref{fig:islands}, the MW and M31 are examples of such islands, not only surrounded by
the sea, but separated from each other by a channel, so they appear not to have been
ionized from outside or to have significantly influenced each other during their reionization. 
For comparison, An example of a reionization map in a scenario  where M31's progenitor partially or completely reionizes the MW progenitor is shown in Fig. 4 of \cite{ocvirk2013}. However, the latter scenario is disfavored because it requires an extreme feedback model, as discussed in \cite{ocvirk2013}.

Investigating further out, we see that the Virgo cluster's progenitor, is also separated from the LG progenitors
by later-to-reionize regions. It is therefore not responsible for the reionization of the LG in the CoDa II simulation.
A similar result was obtained by the CoDa I-AMR simulation with the EMMA code, 
as described in \cite{codadom}, starting from CoDa I initial conditions,
which were an earlier version of CLUES constrained realization initial conditions of the Local Universe.

Finally, the CenA and M81 groups' progenitors also produced their own self-reionized islands. Despite their proximity, they do not appear to have affected the reionization of the local group, or each other, as the respective islands of CenA, M81, MW and M31 are well-separated from each other.

\section{Conclusions}
\label{s:conclusions}

CoDa II (Cosmic Dawn II) is a state-of-the-art, fully-coupled radiation-hydrodynamics 
simulation of reionization and galaxy formation, in a volume large enough to 
model global reionization and with a uniform spatial resolution high enough 
to resolve the formation of all the galactic halos in that volume above $\sim 10^8$\Msun.
Based upon the massively-parallel, hybrid CPU-GPU code RAMSES-CUDATON,
CoDa II utilizes $4096^3$ dark matter particles on a uniform cubic lattice of
$4096^3$ grid cells, which is also used for the gas and radiation field, in a
comoving box $94$ Mpc on a side, centered on the Local Group in a constrained
realization of the Local Universe.  CoDa II is a successor to our first such
simulation, CoDa I, described in O16 and \cite{dawoodbhoy2018},
but with new initial conditions, updated parameters for
the background universe, and a recalibration and modification of our subgrid model
for star formation, chosen to ensure that CoDa II finished reionizing the universe
before $z\sim6$.  For both CoDa I and II, our initial conditions were based 
upon a constrained realization of the Gaussian random density fluctuations 
for the $\Lambda$CDM model, chosen by ``reverse-engineering''
the structure of the present-day Local Universe, 
to yield initial conditions in the linear regime that, when evolved
over cosmic time, will reproduce that observed structure.  This makes CoDa I and II
both suitable for modelling both global and local reionization and its impact
on galaxy formation, including the progenitors of the Local Group.  The galaxy data and background cosmology on which the constrained realization for CoDa II is based were updated and improved relative to those used for CoDa I. 

CoDa II was performed on the hybrid CPU-GPU supercomputer Titan at OLCF, 
using 16384 nodes and 16384 GPUs
(one GPU per node, 4 cores per GPU), for a total of 65536 cores. 
This is 2 times the node and GPU count of CoDa I, and an 8-fold increase in CPU count. 
This is the first time a GPU-accelerated, fully-coupled radiation-hydrodynamics 
galaxy formation code has been used on such a scale.

The simulation accurately describes the properties of the gas and its interaction 
with ionizing radiation, self-consistently, including the growth of typical 
butterfly-shaped ionized regions around the first stars and first galaxies, 
accompanied by photoheating and the subsequent dynamical back-reaction of the gas in
response to pressure-gradients. This back-reaction includes the progressive smoothing of 
small-scale gas structures and the resistance of ionized intergalactic gas to its 
gravitational capture by galactic haloes too small to produce potential
wells deep enough to overcome these pressure forces.

The simulation is in broad agreement with several observational constraints on 
the EoR, such as the Thomson scattering optical depth integrated through the IGM, as
inferred from measurements of fluctuations in the
cosmic microwave background by the \emph{Planck} satellite, 
and the redshift by which reionization ended, as estimated from
measurements of the Lyman alpha opacity of neutral H atoms in the
absorption spectra of high-redshift quasars, due to the intervening IGM.
While the timing of the evolution of the ionizing flux density is reasonable
in CoDa II, its rapid rise at the end of reionization overshoots the observed value 
inferred from observations of quasar absorption spectra at $z=6$ by a factor $\sim 10$. 
This in turn causes the neutral fraction in the post-reionization IGM 
to be too small compared to observations of the Lyman alpha forest at this
redshift, by a similar factor.  This may be a consequence of the existence of additional
small-scale structure in the real universe, responsible for additional H I bound-free 
opacity that limits the mean-free-path of ionizing photons when reionization ends, 
that was missed by the simulation because it was unresolved.

We compute the UV continuum luminosity functions for our simulated galaxy population, 
and find them to be in very good agreement with high-redshift galaxy observations, in particular at $z=6$.
At higher redshifts, a small offset appears, increasing to $\sim 1$ mag at $z=10$.
We apply magnitude cuts to mimic observational detection limits, and find that the cosmic star formation rate associated with haloes 
with $M_{\rm{AB1600}}<-17$ only captures 63\% (18\%) of the total SFR in haloes at $z=6$ ($z=10$). 
Integrating down to fainter magnitudes, $M_{\rm{AB1600}}<-13$, as attempted in deep cluster fields, 
captures a much larger fraction of the total SFR in haloes, 94\% (76\%) at $z=6$ ($z=10$). 
This suggests that calibrating a numerical simulation of galaxy formation by comparing 
its total star formation rate density to observations without applying the magnitude cut
may lead to non-optimal calibration, and in particular too small a simulated star formation rate.

The average star formation rate of individual galaxies increases with their dark matter halo mass, 
yielding a typical SFR=M$_{\rm halo}^\alpha$ relation, when the SFR is globally-averaged,
with a slope $\alpha = 5/3$. At low halo masses (below $\sim 2\times10^9$\Msun), 
the slope becomes steeper, as galaxies become less efficient at forming stars. 
As time goes by and reionization progresses, the average slope for these
low-mass haloes becomes even steeper, 
reflecting the suppression of the SFR in low-mass haloes caused by reionization feedback.

We show unambiguously that this behaviour at low halo mass is the result of the 
spreading, rising ionizing UV background, by comparing our results with a simulation 
without radiative transfer. We note that this radiative reduction of the SFR 
appears to be somewhat less severe than in CoDa I due to different sub-grid physics, but it is still robustly measured.

In contrast, the gas core of high-mass haloes is dense enough to remain cool and/or cool 
down fast enough to keep forming stars, even if in bursts.

Furthermore, we show that environment has a strong impact on the star formation histories 
of low-mass galaxies: galaxies in overdense regions stop forming stars earlier than 
galaxies in underdense regions, supporting the analysis of CoDa I by \cite{dawoodbhoy2018}.

The CoDa II UV luminosity function peaks at ${M_{\rm{AB1600}}}=-11$ at all the studied redshifts, and develops into a plateau at fainter magnitudes, populated by radiation-suppressed, fading galaxies. We provide the full CoDa II UV luminosity functions for 4 epochs in tabular form, across 8 decades in luminosity, from ${M_{\rm{AB1600}}}=-3$ to ${M_{\rm{AB1600}}}=-24$, in the hope that these predictions will be useful for comparison with other simulations, but also to observers analysing high-redshift galaxy observations or designing new deep surveys.
We analyse CoDa II's Local Universe analog region, and find that the Milky Way and its neighbour M31 appear as individual islands in the reionization redshift map. This means that both galaxies reionized in isolation, i.e. they were not reionized by the progenitor of the Virgo cluster, nor by nearby groups, nor by each other.

\section*{Acknowledgements}
This study was performed in the context of several French ANR (Agence Nationale de la Recherche) projects. PO acknowledges support from the French ANR funded project ORAGE (ANR-14-CE33-0016). ND and DA acknowledge funding from the French ANR for project ANR-12-JS05-0001 (EMMA). The CoDa II simulation was performed at Oak Ridge National Laboratory / Oak Ridge Leadership Computing Facility on the Titan supercomputer (INCITE 2016 award AST031). Processing was performed on the Eos and Rhea clusters. Resolution study simulations were performed on Piz Daint at the Swiss National Supercomputing Center (PRACE Tier 0 award, project id pr37). The authors would like to acknowledge the High Performance Computing
center of the University of Strasbourg for supporting this work by
providing scientific support and access to computing resources. Part of
the computing resources were funded by the Equipex Equip\@Meso project
(Programme Investissements d'Avenir) and the CPER Alsacalcul/Big Data.

The CoDaII-DM2048 / ESMDPL\_2048 simulation has been performed at LRZ Munich within the project pr74no.

ITI was supported by the Science and Technology Facilities Council [grant number ST/L000652/1]. JS acknowledges support from the ``l'Or\'eal-UNESCO Pour les femmes et la Science'' and the ``Centre National d'\'Etudes Spatiales (CNES)'' postdoctoral fellowship programs.
KA was supported by NRF-2016R1D1A1B04935414.
YH has been partially supported by the Israel Science Foundation (1013/12). 

GY also acknowledges support from MINECO-FEDER under research grants AYA2012-31101, AYA2015-63810-P and AYA2015-63810-P, as well as MICIU/FEDER under grant PGC2018-094975-C21.

PRS was supported in part by U.S. NSF grant AST-1009799, NASA grant NNX11AE09G, NASA/JPL grant RSA Nos. 1492788 and 1515294, and supercomputer resources from NSF XSEDE grant TG-AST090005 and the Texas Advanced Computing Center (TACC) at the University of Texas at Austin. PO thanks Y. Dubois, F. Roy and Y. Rasera for their precious help dealing with SN feedback in RAMSES and various hacks in pFoF. This work made use of v2.1 of the Binary Population and Spectral Synthesis (BPASS)
models as last described in \cite{bpass21}.

\bibliographystyle{mn2e}
\bibliography{mybib}

\appendix

\section{Resolution study on suppression of star formation}
\label{s:resstudy}
\begin{figure*}
    \centering
    {\includegraphics[width=0.5075\linewidth,clip]{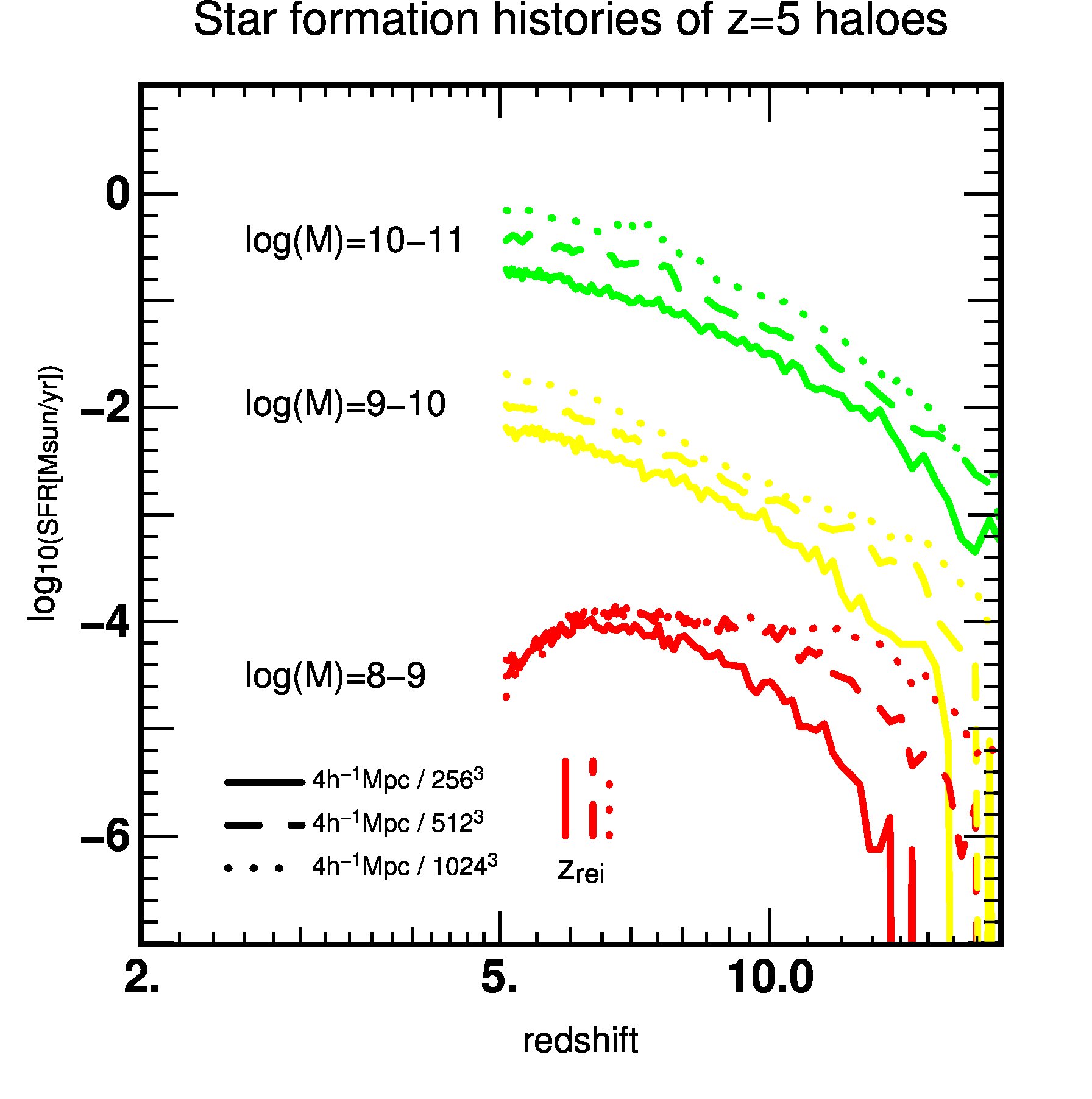}}
    {\includegraphics[width=0.47\linewidth,clip]{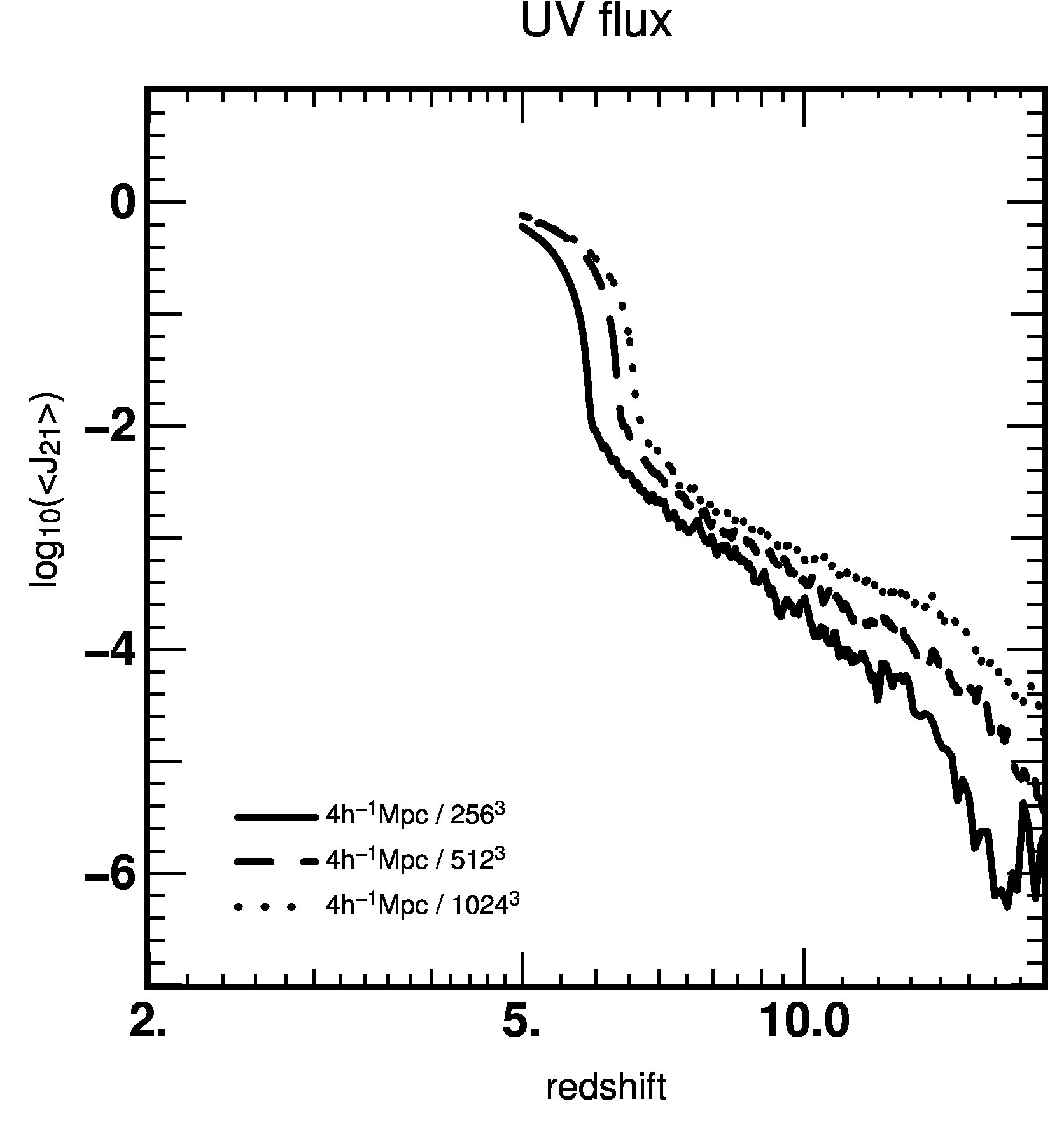}}
  \caption{Results of the resolution study simulations with full CoDaII physics listed in Table A1. The test box is 4 \chmpc on a side, at different spatial resolutions. {\em Solid line}: fiducial CoDa II resolution (256$^3$), {\em dashed line}: $512^3$, {\em dotted line}: $1024^3$. {\em Left:} Average star formation histories for 3 halo mass bins. The vertical lines above the "zrei" label indicate the reionization redshift of each simulation. The mass bins correspond to the haloes final mass (i.e. halo mass measured in the final simulation timestep, at z=5. {\rm Right:} Volume-averaged ionizing UV flux J$_{21}$ for these simulations.}
    \label{f:res_study}
\end{figure*}

We conducted a resolution study to determine the impact of spatial resolution on the suppression of star formation in low mass haloes due to radiative feedback. To this end we performed a series of 3 simulations of a test box 4 \chmpc on a side, with the full CoDa II physics, using exactly the same parameters as given in Table 1, except for grid cell size and dark matter particle mass, which therefore impact spatial resolution and mass resolution. We explored spatial (mass)  resolutions 2 (8) times and 4 (64) times higher than the fiducial CoDa II setup. These simulations are described in Table \ref{t:resstudy}.
The average star formation histories of z=5 haloes were then computed for 3 different bins of dark matter halo mass and are shown in the left panel of Fig. \ref{f:res_study}. The global shapes of the star formation histories for each mass bin are similar to those of Fig. \ref{f:fbstudy} in the full physics case ("SN, RT"). The highest mass bin of Fig. \ref{f:fbstudy} is missing in this resolution study because it was performed in a 4 \chmpc instead of 8 \chmpc box, to keep the computational cost of the study in check. This is not a problem, however, since it is in low mass haloes that suppression is strongest and matters the most. That is the focus of our resolution study.

The left panel of Fig. \ref{f:res_study} shows that for the two most massive bins, i.e. between $10^9$ and $10^{11}$ \Msun, where the suppression is almost non-existent, increasing spatial resolution by a factor of 2 yields an increase of star formation rate of a factor $\sim 2$. A similar increase is also seen at early times in the $10^{8-9}$ \Msun halo mass bin, where star formation rises earlier in the high resolution runs. However, in all 3 runs, the low mass haloes experience a star formation plateau at a similar level of $\sim 10^{-4}$ \Msun/yr. Reionization ends for all 3 simulations at  redshifts in the range z=6-7 (shown by the vertical solid, dashed and dotted lines), and the post-reionization epochs are marked, as in Fig. \ref{f:res_study}, by a decline of SFR in the low mass haloes. The decrease is seen at all resolutions, and the timing and slope of the suppression are similar across all 3 runs.

\begin{figure*}
    \centering
    {\includegraphics[width=0.48\linewidth,clip]{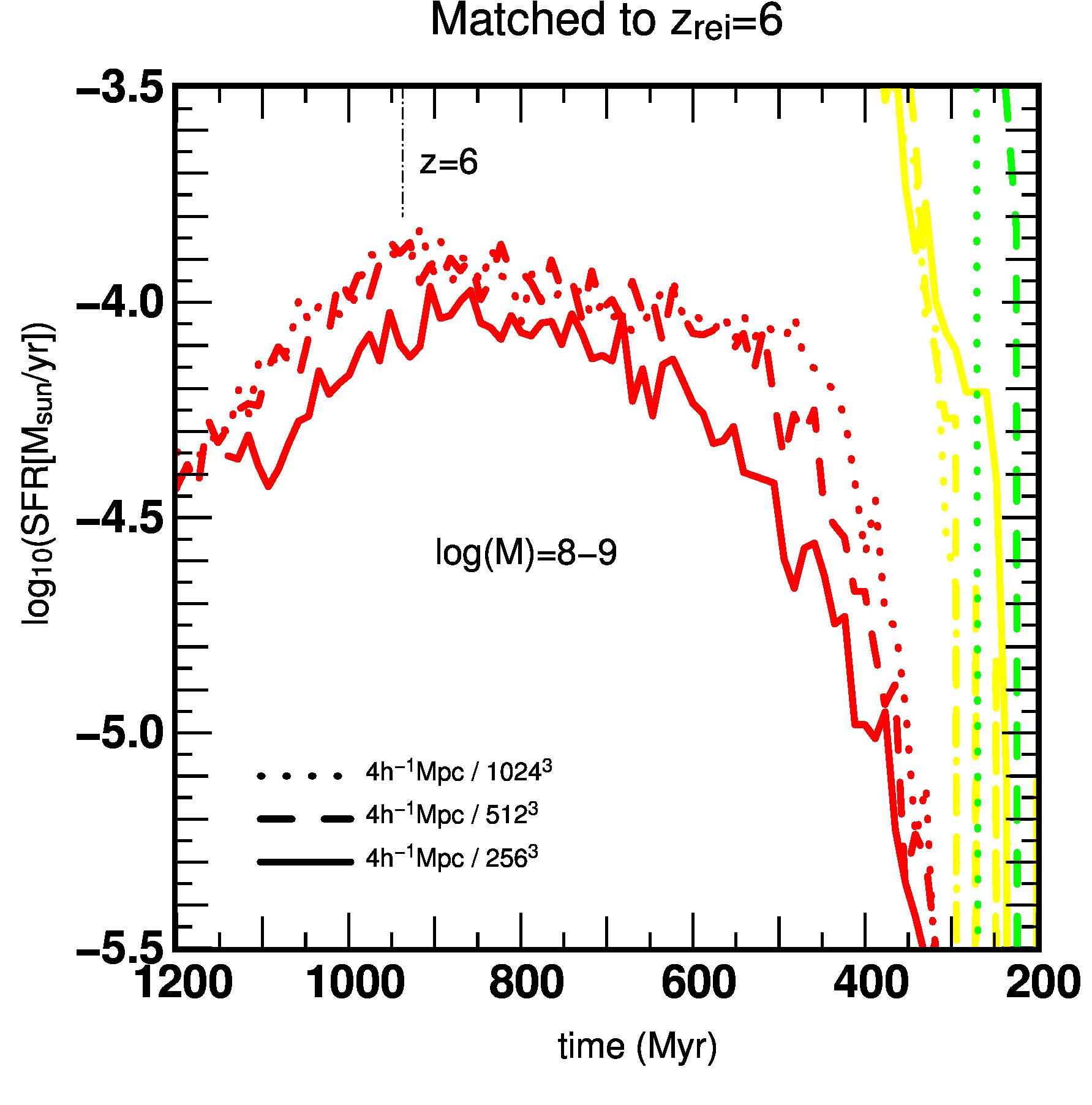}}
    {\includegraphics[width=0.47\linewidth,clip]{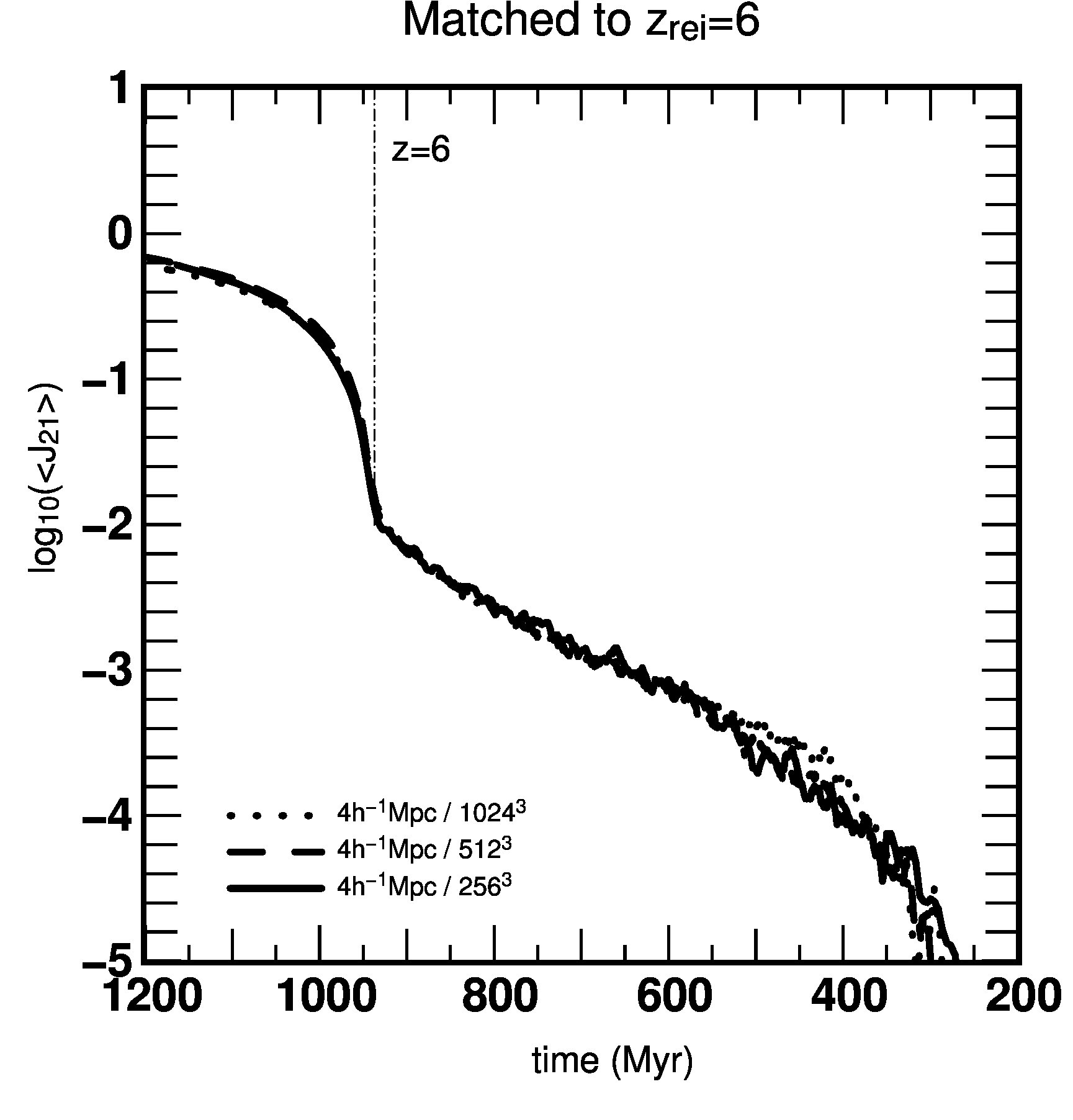}}
    \caption{Same as Fig. A1, but the simulations have been shifted in time to align them, so that all 3 runs have $z_{\rm rei}=6$. This results in a shift of -17 Myr, 66 Myr and 112 Myr for the 256$^3$, $512^3$ and $1024^3$ run, respectively. The x axis now shows time since Big Bang, with z=6 marked as a vertical dot-dashed line for reference. The left panel is zoomed on the low mass bin.}
      
    \label{f:res_study_zshift}
\end{figure*}

{
To account for the slight differences in reionization redshifts between the 3 runs, we shifted the runs {\em in time} to align their reionization histories, so that they have the same reionization redshift z=6. 
The resulting average halo star formation histories and J$_{21}$ are shown in Fig. \ref{f:res_study_zshift}.

With these small time shifts applied to each, the curves of J$_{21}$ lie virtually on top of each other, especially after $\sim 550$ Myr, demonstrating that the 3 runs have very similar levels of irradiation at each stage of their respective reionization histories.

We can now take a closer look at the SFRs of the low mass bin, as shown in the zoomed left panel of Fig. \ref{f:res_study_zshift}. The slightly earlier onset and rise of SFR at higher resolutions is confirmed, as is the quasi-plateau at a SFR=$10^{-4}$ \Msun/yr, from 600 Myr to z=6, shared by all 3 resolution runs. During this plateau, the SFRs of all 3 runs are very consistent, within 0.1 dex of each other. The same sharp decline of SFR takes place at z=6 for all 3 runs.
}



We already explored the impact of numerical resolution on star formation suppression in CoDa I in \cite{codaI}, and found little impact when increasing resolution by a factor 2. The new results of this appendix confirm that, despite the revised subgrid model for star formation for CoDaII, which results in a milder suppression than CoDa I, suppression is still seen in $10^{8-9}$ \Msun haloes, even at a resolution 4 times higher.

Some other efforts have also been made to address the question of radiative suppression of star formation in low mass haloes based upon cosmological hydrodynamical simulations with RT.  Although their methodology and metrics for measuring suppression may differ from ours, there is reasonable agreement with at least some of them. For instance, \cite{Wu2019_2} is particularly valuable because of the use of a completely different code, AREPO-RT. They found that photo-heating reduces the SFR by 50\% in haloes below 10$^{8.4}$ \Msun at z=6 and 10$^{8.8}$ \Msun at z=5. At both redshifts, their suppression mass is within the 10$^{8-9}$ \Msun mass bin where suppression can be seen to take place in Fig. \ref{f:res_study}.

A similar agreement is found with \cite{pawlik2015}, also using a completely different methodology (Smoothed Particles Hydrodynamics and a form of adaptive ray-tracing). They found a break at the low-mass end of the SFR-halo mass relation, which evolves with redshift.
Their simulations reionize earlier than CoDa II, at $z\sim 8$. At this redshift, they find a break in the low-mass end of the SFR-halo mass relation, at $5.10^9$ \Msun, characteristic of suppression of star formation by photo-heating and supernova feedback. Altough this mass increases with decreasing redshift, the overlap between the halo mass range where suppression takes place in CoDa II and \cite{pawlik2015} is encouraging.

The rough agreement between these examples from the literature and our own results, along with the resolution study presented here, suggests that the suppression in star formation seen in CoDa II is not an artifact due to numerical resolution. 

\begin{table*}
  \label{t:resstudy}  
  \begin{center}
    \begin{tabular}{lccccc}
      \hline
    Simulation & N$_{\rm DM}$ & M$_{\rm DM}$ (\Msun) & dx (ckpc) & dx$_{\rm phys}$(kpc,z=5)  &Comment \\
    \hline
    L04N0256  &  $256^3$ & $4.07 \times 10^5$ & 23.06 & 3.84 & Fiducial \\
    L04N0512  & $512^3$ & $6.21 \times 10^4$ & 11.53 & 1.92   &dx$_{\rm fiducial}$/2 \\
    L04N1024  & $1024^3$ & $7.77 \times 10^3$ & 5.76 & 0.96   & dx$_{\rm fiducial}$/4 \\
    \hline
  \end{tabular}
  \caption{Summary table of the simulations of our resolution study, giving the number of particles, the dark matter particle mass, and the spatial resolution in comoving and physical physical kpc at z=5. All simulations are performed with full CoDa II physics.}
  \end{center}
\end{table*}

\section{Luminosity functions}
We give the CoDa II luminosity functions at 4 epochs in Table \ref{t:LFs}, in the hope that they may be of use to the community and in the design, analysis and interpretation of future surveys.

\begin{table*}
\begin{tabular}{c|cccc|cccc|cccc|cccc} 
M$_{ \rm AB1600}$  & \multicolumn{16}{c}{ $\log_{10}(\Phi[{\rm Mag^{-1} Mpc^{-3} }]$  median (1),  average (2),  min (3),  max (4))    } \\ 
 \hline 
   & \multicolumn{4}{c|}{z=6} & \multicolumn{4}{c|}{z=7} & \multicolumn{4}{c|}{z=8}  &  \multicolumn{4}{c}{z=10}  \\ 
  &   (1) & (2) &(3) &(4)  &  (1) & (2) &(3) &(4)      &  (1) & (2) &(3) &(4)  &  (1) & (2) &(3) &(4) \\  \hline 
-24.0 &   &   &   &   &   &   &   &   &   &   &   &   &   &   &   &   \\ 
-23.5 &   & -5.58 &   & -4.88 &   &   &   &   &   &   &   &   &   &   &   &   \\ 
-23.0 &   & -5.58 &   & -4.88 &   & -5.58 &   & -4.88 &   &   &   &   &   &   &   &   \\ 
-22.5 & -4.88 & -4.80 &   & -4.58 &   & -5.28 &   & -4.88 &   &   &   &   &   &   &   &   \\ 
-22.0 & -4.41 & -4.41 &   & -4.11 & -4.88 & -4.98 &   & -4.58 &   & -5.28 &   & -4.88 &   &   &   &   \\ 
-21.5 & -4.18 & -4.17 & -4.41 & -3.93 & -4.88 & -4.74 &   & -4.41 &   & -5.28 &   & -4.58 &   &   &   &   \\ 
-21.0 & -3.74 & -3.80 & -4.18 & -3.65 & -4.28 & -4.26 & -4.58 & -4.04 & -4.88 & -4.74 &   & -4.41 &   &   &   &   \\ 
-20.5 & -3.44 & -3.47 & -3.74 & -3.28 & -3.98 & -4.04 & -4.88 & -3.80 & -4.88 & -4.50 & -4.88 & -4.18 &   &   &   &   \\ 
-20.0 & -3.18 & -3.21 & -3.50 & -3.06 & -3.60 & -3.62 & -3.88 & -3.50 & -4.28 & -4.17 & -4.41 & -3.98 &   & -5.28 &   & -4.58 \\ 
-19.5 & -2.95 & -3.00 & -3.36 & -2.87 & -3.38 & -3.38 & -3.65 & -3.25 & -3.98 & -3.89 & -4.11 & -3.71 & -4.88 & -4.88 &   & -4.58 \\ 
-19.0 & -2.67 & -2.70 & -2.86 & -2.55 & -3.04 & -3.07 & -3.38 & -2.90 & -3.49 & -3.50 & -3.93 & -3.35 & -4.41 & -4.44 & -4.58 & -4.41 \\ 
-18.5 & -2.45 & -2.50 & -2.73 & -2.39 & -2.74 & -2.77 & -3.00 & -2.67 & -3.16 & -3.17 & -3.39 & -3.06 & -4.28 & -4.24 & -4.58 & -4.11 \\ 
-18.0 & -2.32 & -2.33 & -2.51 & -2.22 & -2.59 & -2.61 & -2.73 & -2.49 & -2.85 & -2.89 & -3.08 & -2.81 & -3.88 & -3.90 & -4.11 & -3.74 \\ 
-17.5 & -2.13 & -2.15 & -2.31 & -2.07 & -2.38 & -2.41 & -2.60 & -2.30 & -2.74 & -2.73 & -2.89 & -2.60 & -3.71 & -3.60 & -3.77 & -3.44 \\ 
-17.0 & -2.01 & -1.98 & -2.16 & -1.87 & -2.23 & -2.22 & -2.37 & -2.11 & -2.51 & -2.51 & -2.64 & -2.43 & -3.27 & -3.25 & -3.47 & -3.15 \\ 
-16.5 & -1.84 & -1.83 & -1.98 & -1.73 & -2.05 & -2.05 & -2.22 & -1.95 & -2.31 & -2.32 & -2.48 & -2.20 & -3.00 & -3.02 & -3.30 & -2.87 \\ 
-16.0 & -1.68 & -1.68 & -1.84 & -1.59 & -1.88 & -1.88 & -2.03 & -1.78 & -2.16 & -2.13 & -2.31 & -2.01 & -2.81 & -2.81 & -2.93 & -2.72 \\ 
-15.5 & -1.55 & -1.53 & -1.69 & -1.44 & -1.74 & -1.73 & -1.89 & -1.62 & -1.95 & -1.95 & -2.11 & -1.84 & -2.54 & -2.57 & -2.70 & -2.45 \\ 
-15.0 & -1.42 & -1.41 & -1.55 & -1.32 & -1.57 & -1.57 & -1.71 & -1.48 & -1.81 & -1.79 & -1.93 & -1.69 & -2.41 & -2.37 & -2.56 & -2.24 \\ 
-14.5 & -1.30 & -1.30 & -1.41 & -1.20 & -1.45 & -1.43 & -1.57 & -1.34 & -1.63 & -1.62 & -1.79 & -1.53 & -2.20 & -2.16 & -2.32 & -2.05 \\ 
-14.0 & -1.18 & -1.18 & -1.30 & -1.10 & -1.32 & -1.31 & -1.43 & -1.22 & -1.50 & -1.47 & -1.59 & -1.38 & -2.01 & -1.96 & -2.11 & -1.86 \\ 
-13.5 & -1.08 & -1.08 & -1.19 & -0.99 & -1.19 & -1.17 & -1.29 & -1.08 & -1.35 & -1.34 & -1.45 & -1.24 & -1.82 & -1.80 & -1.97 & -1.68 \\ 
-13.0 & -0.98 & -0.98 & -1.08 & -0.89 & -1.07 & -1.06 & -1.17 & -0.97 & -1.21 & -1.19 & -1.31 & -1.10 & -1.65 & -1.62 & -1.75 & -1.51 \\ 
-12.5 & -0.90 & -0.89 & -0.98 & -0.81 & -0.96 & -0.94 & -1.05 & -0.84 & -1.09 & -1.06 & -1.18 & -0.96 & -1.50 & -1.46 & -1.58 & -1.34 \\ 
-12.0 & -0.84 & -0.82 & -0.88 & -0.74 & -0.87 & -0.84 & -0.93 & -0.75 & -1.00 & -0.95 & -1.06 & -0.84 & -1.39 & -1.33 & -1.43 & -1.20 \\ 
-11.5 & -0.81 & -0.78 & -0.83 & -0.70 & -0.81 & -0.78 & -0.86 & -0.67 & -0.92 & -0.87 & -0.96 & -0.75 & -1.31 & -1.24 & -1.33 & -1.11 \\ 
-11.0 & -0.79 & -0.76 & -0.80 & -0.69 & -0.77 & -0.72 & -0.79 & -0.61 & -0.88 & -0.81 & -0.89 & -0.68 & -1.24 & -1.16 & -1.26 & -1.02 \\ 
-10.5 & -0.88 & -0.85 & -0.89 & -0.77 & -0.87 & -0.81 & -0.88 & -0.70 & -0.97 & -0.90 & -0.97 & -0.77 & -1.32 & -1.25 & -1.34 & -1.09 \\ 
-10.0 & -0.95 & -0.93 & -0.97 & -0.85 & -0.95 & -0.90 & -0.97 & -0.78 & -1.07 & -1.01 & -1.10 & -0.87 & -1.48 & -1.41 & -1.53 & -1.24 \\ 
-9.5 & -0.93 & -0.92 & -0.97 & -0.84 & -0.96 & -0.91 & -1.00 & -0.79 & -1.10 & -1.03 & -1.13 & -0.88 & -1.51 & -1.45 & -1.58 & -1.28 \\ 
-9.0 & -0.97 & -0.96 & -1.02 & -0.86 & -1.01 & -0.97 & -1.07 & -0.83 & -1.15 & -1.10 & -1.21 & -0.94 & -1.63 & -1.56 & -1.70 & -1.38 \\ 
-8.5 & -0.95 & -0.94 & -1.01 & -0.83 & -1.00 & -0.97 & -1.08 & -0.83 & -1.17 & -1.13 & -1.26 & -0.97 & -1.70 & -1.63 & -1.78 & -1.44 \\ 
-8.0 & -0.95 & -0.93 & -1.01 & -0.82 & -1.05 & -1.02 & -1.14 & -0.86 & -1.22 & -1.18 & -1.32 & -1.02 & -1.79 & -1.74 & -1.92 & -1.54 \\ 
-7.5 & -0.97 & -0.96 & -1.08 & -0.83 & -1.14 & -1.11 & -1.25 & -0.94 & -1.35 & -1.32 & -1.48 & -1.14 & -2.01 & -1.97 & -2.17 & -1.76 \\ 
-7.0 & -1.04 & -1.03 & -1.16 & -0.88 & -1.26 & -1.24 & -1.39 & -1.07 & -1.55 & -1.52 & -1.71 & -1.32 & -2.32 & -2.28 & -2.55 & -2.03 \\ 
-6.5 & -1.14 & -1.12 & -1.28 & -0.95 & -1.45 & -1.42 & -1.60 & -1.23 & -1.80 & -1.76 & -1.96 & -1.55 & -2.90 & -2.81 & -3.05 & -2.51 \\ 
-6.0 & -1.16 & -1.12 & -1.26 & -0.95 & -1.56 & -1.51 & -1.66 & -1.32 & -2.09 & -2.03 & -2.21 & -1.81 & -3.84 & -3.74 & -4.18 & -3.47 \\ 
-5.5 & -1.56 & -1.49 & -1.65 & -1.28 & -2.50 & -2.43 & -2.65 & -2.18 & -3.60 & -3.60 & -3.84 & -3.39 &   &   &   &   \\ 
-5.0 & -3.23 & -3.08 & -3.45 & -2.76 &   &   &   &   &   &   &   &   &   &   &   &   \\ 
-4.5 &   &   &   &   &   &   &   &   &   &   &   &   &   &   &   &   \\ 
-4.0 &   &   &   &   &   &   &   &   &   &   &   &   &   &   &   &   \\ 
-3.5 &   &   &   &   &   &   &   &   &   &   &   &   &   &   &   &   \\ 
-3.0 &   &   &   &   &   &   &   &   &   &   &   &   &   &   &   &   \\ 
\end{tabular}
\caption{CoDa II UV luminosity functions at 4 different epochs. Each group of 4 columns corresponds to a different redshift. Within each group, the 4 columns give the base-10 logarithm of (1): the median LF, (2): the average LF, (3): the minimum LF, (4): the maximum LF. The LF statistics have been determined using 5 independent sub-boxes each spanning 1/5 of the total CoDa II volume.}
\label{t:LFs}
\end{table*}

\section{Stars beyond $R_{200}$}
\label{s:orphanstars}
\begin{figure*}
  \label{f:orphanstars}
  \centering
  \begin{tabular}{cc}
    {\includegraphics[width=0.39\linewidth,clip]{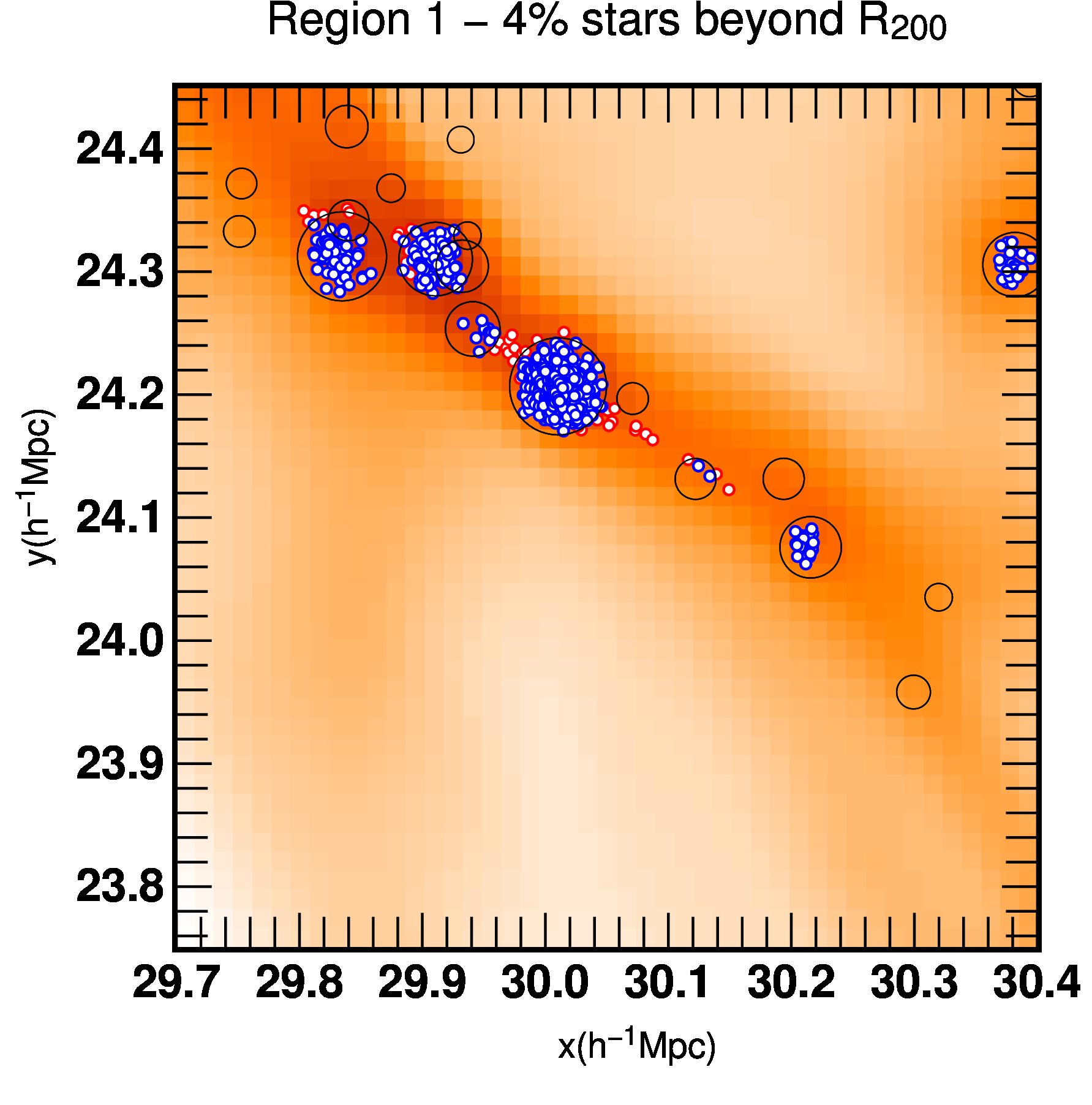}}&
    {\includegraphics[width=0.39\linewidth,clip]{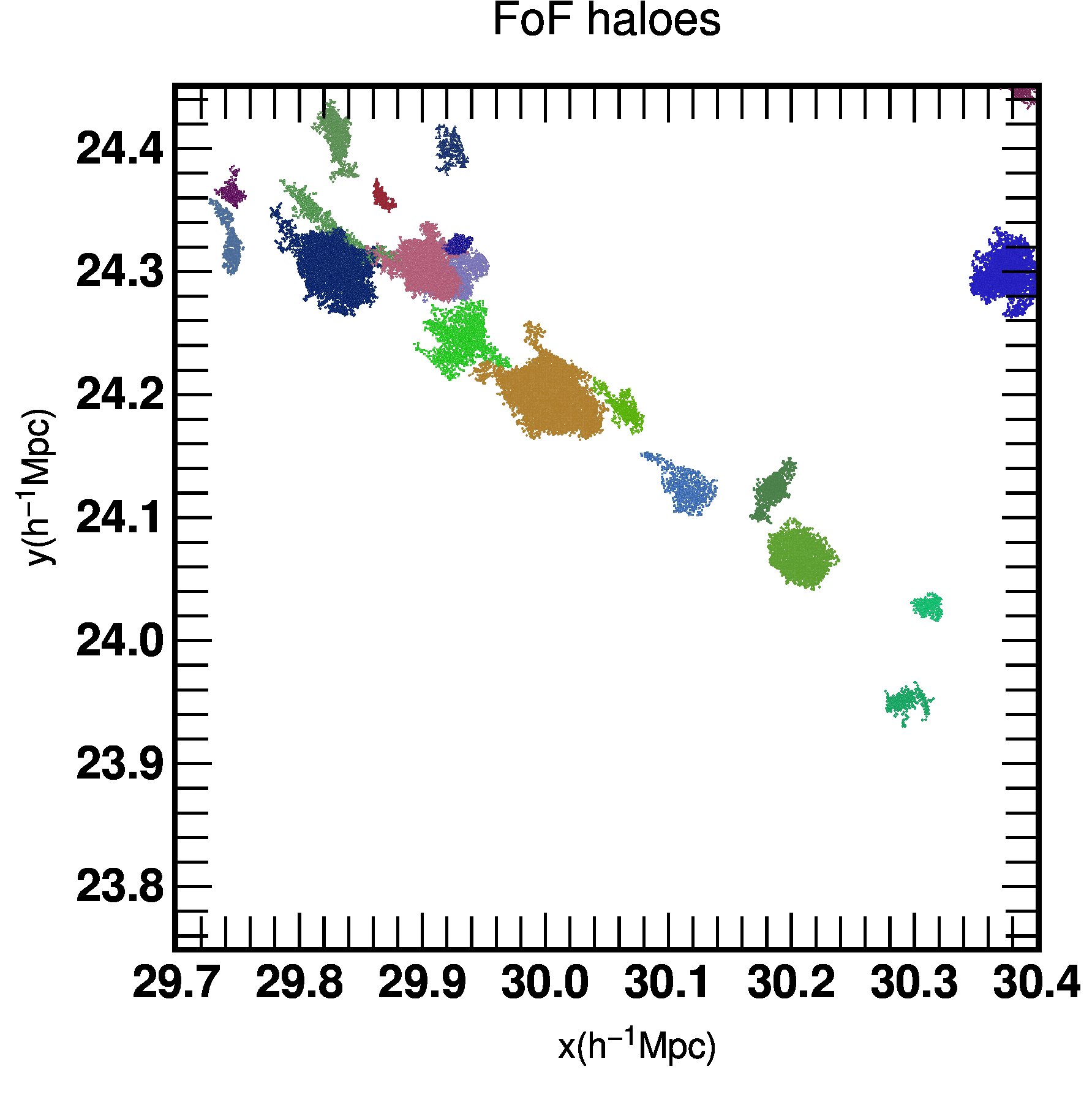}} \\
    {\includegraphics[width=0.39\linewidth,clip]{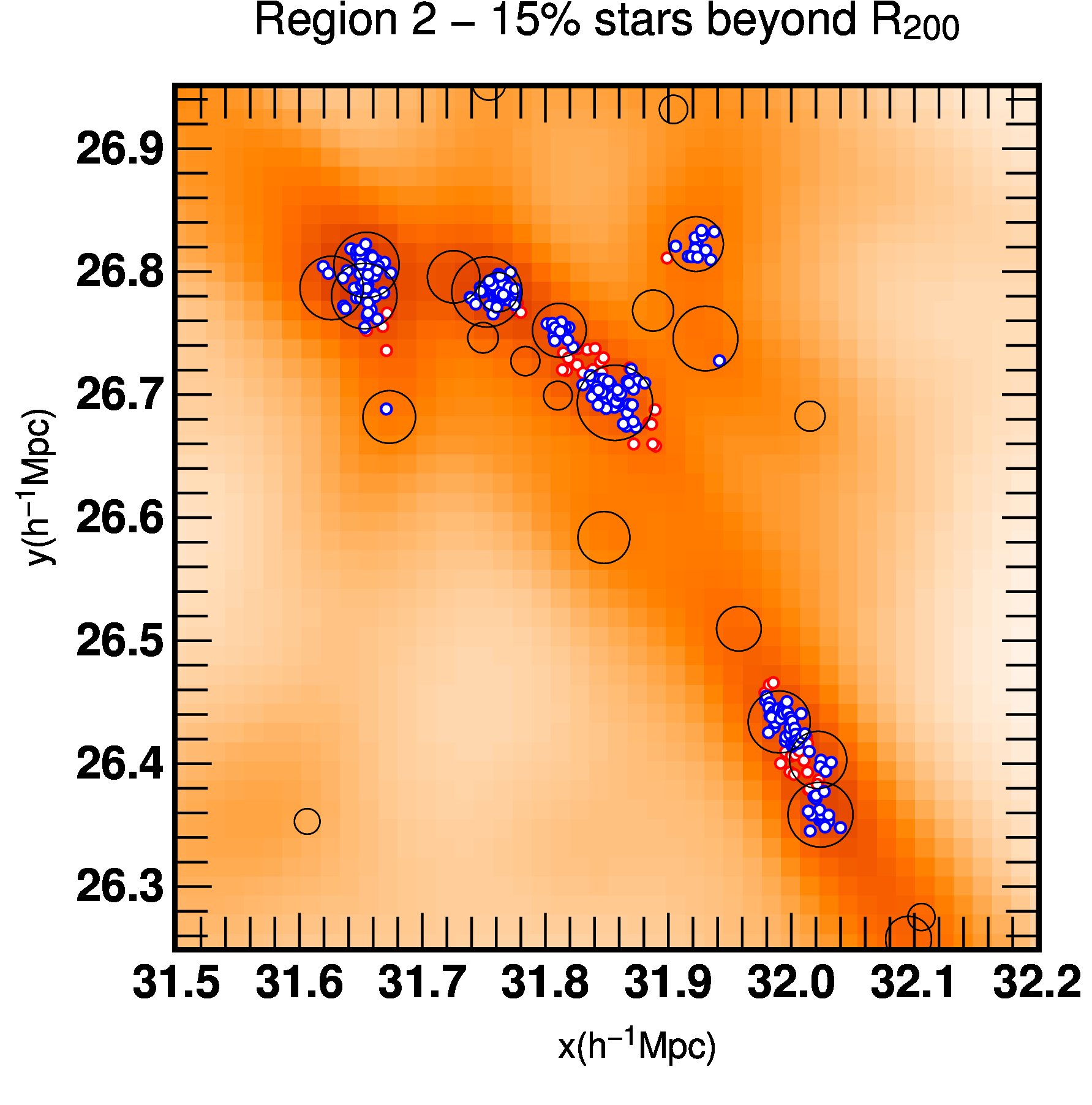}}&
    {\includegraphics[width=0.39\linewidth,clip]{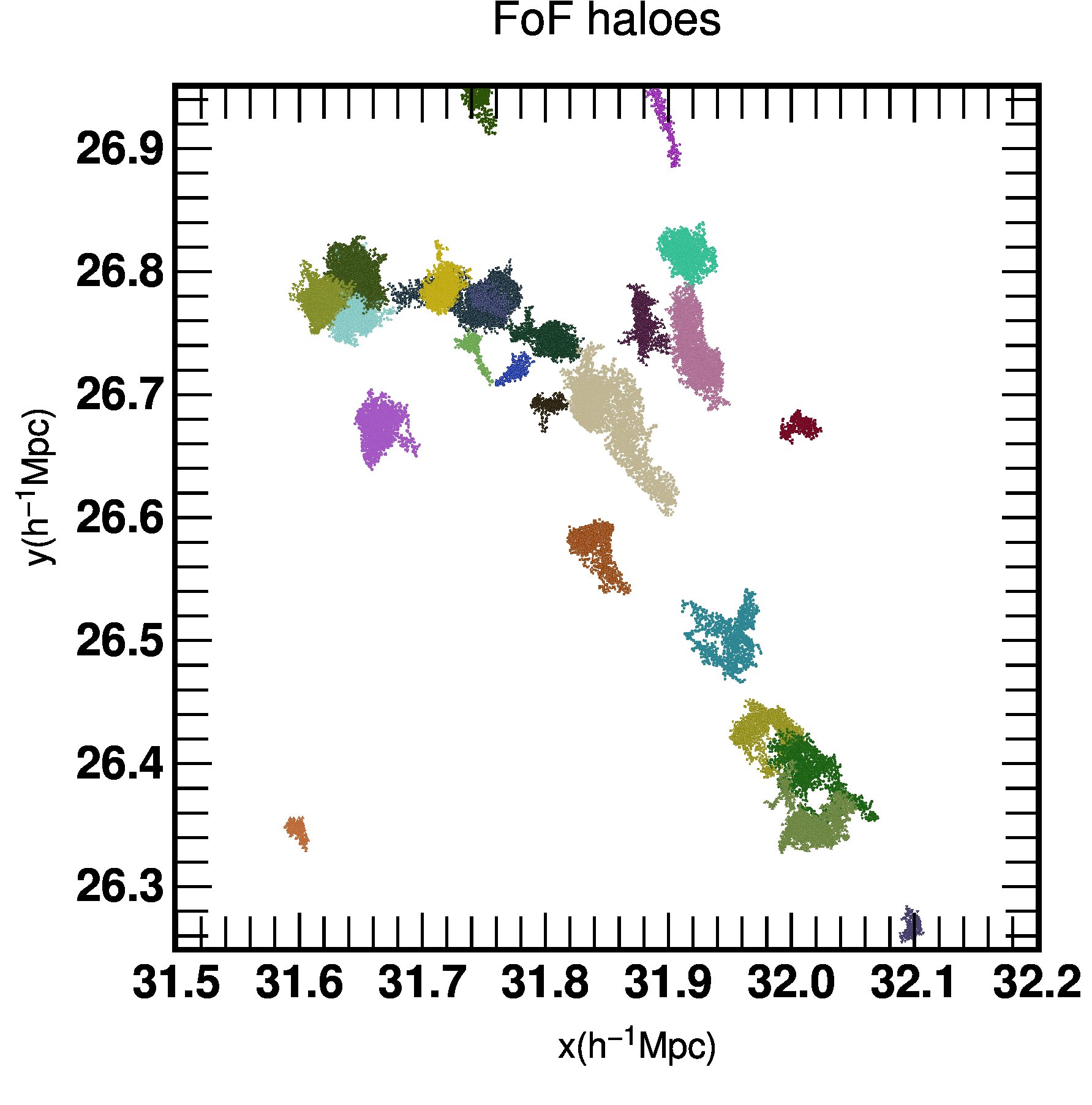}} \\
    {\includegraphics[width=0.39\linewidth,clip]{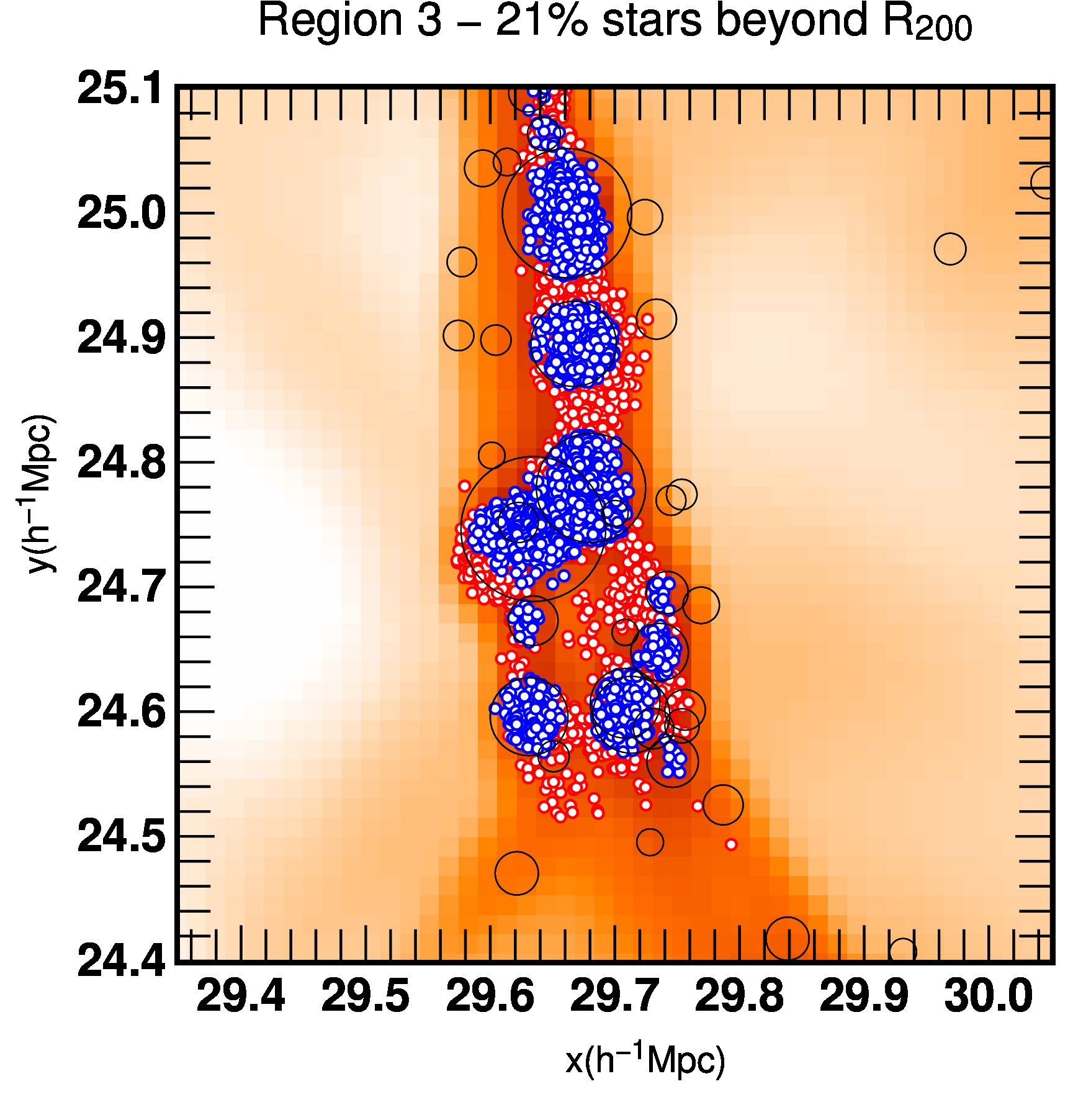}}&
    {\includegraphics[width=0.39\linewidth,clip]{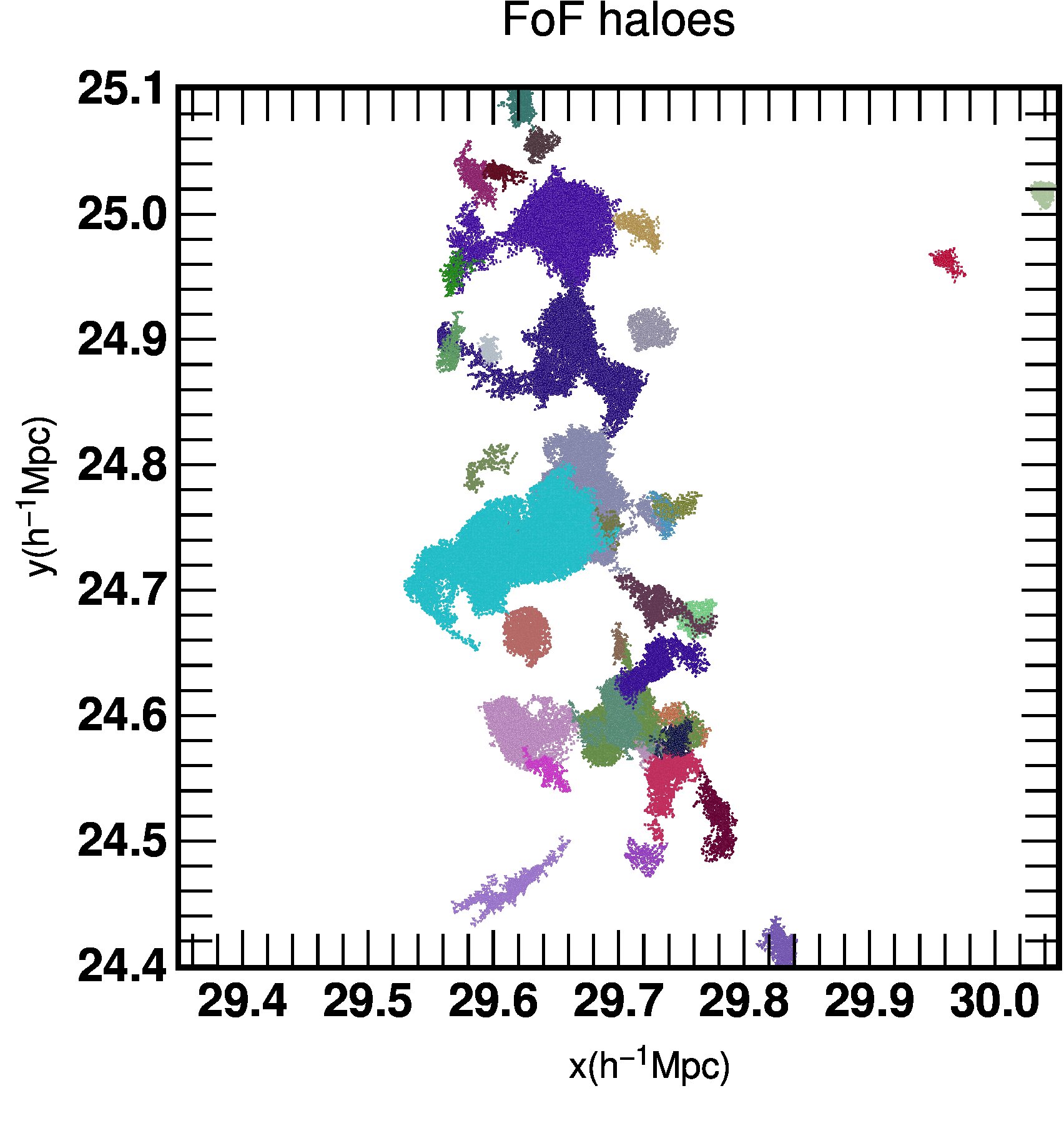}} \\
  \end{tabular}
  \caption{{\em Left:} Maps of the gas density for 3 different example regions in CoDa II, overplotted with the $R_{200}$ spheres of dark matter halos more massive than M$>10^8$ \Msun (black circles). Also shown are the stellar particles, in blue if the particle is  inside of the $R_{200}$ sphere of a dark matter halo and red otherwise. The percentage of stars beyond $R_{200}$ is indicated in the title of each panel. The most massive halo in the top, middle and bottom panel is 5.5x$10^9$, 2.6x$10^9$ and 1.8x$10^{10}$ \Msun, respectively. {\em Right:} Maps of the dark matter particles of the FoF haloes of the corresponding left panel. A color is chosen randomly for each halo.
    }
\end{figure*}

{ 
Following our finding in Sec. \ref{s:csfrd} that a fraction of the stars are found beyond $R_{200}$, we provide a few maps to illustrate this aspect and gain insight into the relative distributions of stars and dark matter haloes in CoDa II.


The left column of Fig. \ref{f:orphanstars} shows a gas density map overplotted with the $R_{200}$ spheres of dark matter halos more massive than M$>10^8$ \Msun (black circles), along with stars, within and beyond $R_{200}$, for 3 typical star-forming regions of the simulation. The slice is 30 cells thick, i.e. about $470 {\rm h^{-1} ckpc}$. The fraction of stars beyond $R_{200}$ increases from top to bottom, and is given in the title of each panel. In most cases, star particles are contained within the $R_{200}$ of their parent haloes (blue circles), and the latter are well aligned along gas filaments, forming strings of haloes. Stars beyond $R_{200}$ are mostly located between pairs of neighbour haloes within these strings. This is confirmed in the bottom plot, which contains a higher density region and more massive haloes, and where the fraction of stars beyond $R_{200}$ reaches 21\%, comparable to what we found globally in Sec. \ref{s:csfrd} at z=6. 
Here we show that the fraction of stars beyond $R_{200}$ is primarily the result of approximating FoF halo shapes by a sphere. The right column of Fig. \ref{f:orphanstars} shows the particles of the FoF dark matter haloes of each region. They are clearly not spherical, always triaxial, often elongated in the direction of the filament, and often feature bridge-like structures linking them. Stars beyond $R_{200}$ fall along these aspherical features. At higher redshifts, halo shapes become even more aspherical, accentuating this effect and increasing the number of stars found beyond $R_{200}$.
If one considers more complex halo shapes (using convex hulls for instance)  one would expect that the census of stars within haloes would be close to complete. Still a small fraction of stars might be formed in small halos below the assumed minimum halo mass.
}

\bsp	
\label{lastpage}
\end{document}